**THE EUROPEAN PHYSICAL JOURNAL C**

# Optical images of the Kerr–Sen black hole and thin accretion disk

Pei Wang[1], Sen Guo[1,a], Li-Fang Li[2,b], Zhan-Feng Mai[3], Bo-Feng Wu[3], Wen-Hao Deng[4], Qing-Quan Jiang[4]

[1] College of Physics and Electronic Engineering, Chongqing Normal University, Chongqing 401331, People's Republic of China
[2] Center for Gravitational Wave Experiment, National Microgravity Laboratory, Institute of Mechanics, Chinese Academy of Sciences, Beijing 100190, People's Republic of China
[3] Guangxi Key Laboratory for Relativistic Astrophysics, School of Physical Science and Technology, Guangxi University, Nanning 530004, People's Republic of China
[4] School of Physics and Astronomy, China West Normal University, Nanchong 637009, People's Republic of China

**Abstract** This paper investigates the observable properties of a Kerr–Sen black hole surrounded by a thin accretion disk, focusing on the impact of the black hole's spin and charge on the image. Using ray-tracing techniques, we conduct a detailed analysis of the black hole's image, redshift distribution, and intensity distributions at different observation frequencies. The results demonstrate that spin has a more significant effect on the distortion of the inner shadow than charge, and the observer's inclination angle plays a critical role in shaping the redshift distribution, especially near the innermost stable circular orbit. Additionally, the intensity is found to be higher at 86 GHz than at 230 GHz. This study highlights the crucial role of the accretion disk's geometry in determining the black hole's image and redshift effects, thereby providing a refined theoretical framework to guide future observational efforts targeting the Kerr–Sen black hole and its electromagnetic signals.

## 1 introduction

General relativity (GR) provides the foundational framework for black hole physics and cosmology, predicting the existence of black holes. Initially, black holes were regarded as merely theoretical solutions to Einstein's field equations. However, over the past few decades, a substantial body of observational data has steadily accumulated, offering compelling evidence for their astrophysical reality [1,2]. A landmark event in this field occurred in 2019, when the Event Horizon Telescope (EHT) collaboration revealed the first direct image of a black hole shadow, specifically that of M87*

[3]. This breakthrough constituted the first direct observational confirmation of GR's predictions in the strong-field regime. Building on this achievement, the EHT collaboration released a polarimetric image of M87* in 2021, unveiling the complex magnetic field structure surrounding the black hole and providing deeper insights into the mechanisms responsible for relativistic jet formation [4,5]. In 2022, the EHT presented the image of Sagittarius A* (Sgr A*), the supermassive black hole at the center of the Milky Way. This observation not only provided direct visual confirmation of a supermassive black hole within our Galaxy, but also corroborated prior constraints derived from stellar orbital dynamics, further validating the applicability of GR in describing astrophysical black holes [6]. These groundbreaking observations represent a significant advancement in the study of strong gravitational fields and continue to stimulate progress in both theoretical modeling and high-resolution observational techniques in the field of black hole physics [7].

The shape and size of the shadows of M87* and Sgr A* have played a crucial role in inferring their rotational characteristics, thereby strengthening support for the Kerr black hole model [4,5,8–10]. These observations provide key insights into the phenomenology of strong-field gravity, particularly in the context of gravitational lensing. The black hole shadow, delineated by the innermost unstable photon orbit, manifests as a central dark region, while the surrounding bright ring corresponds to the outer photon orbit [11–13]. Both the shadow and the photon ring serve as invaluable tools for estimating critical parameters of the black hole, including its mass, spin, magnetic field, and the properties of its accretion disk [14,15]. The structure of the photon ring is strongly modulated by gravitational lensing and the emission from the accretion disk, with the disk's geometry and radiation playing a decisive role in determining the observed brightness [13].

The intrinsic characteristics of the black hole shadow are influenced by the underlying spacetime geometry, whereas the surrounding bright photon ring region is primarily affected by the radiative properties and structure of the accretion disk. Previous theoretical studies have classified black hole shadows and photon rings, emphasizing the pivotal influence of accretion disk microphysics [16,17]. The introduction of the "critical curve," which coincides with the photon orbit, has proven essential in delineating the black hole shadow and has stimulated a substantial body of follow-up investigations. Accretion disks remain a central focus in astronomical observations, as they play a crucial role in shaping black hole images [18,19]. Although simplified models of accretion disks may not fully capture the complexities of astrophysical environments, they nonetheless yield valuable theoretical predictions for the interpretation of high-resolution images [20,21].

Simulated images of Kerr black holes surrounded by optically thin accretion disks reveal two distinct features: a central brightness depression and a narrow, bright "photon ring" [22–24]. Building upon this framework, Wang et al. explored Kerr–de Sitter black holes, examining the influence of a nonzero cosmological constant on the resulting image characteristics [25–27]. Hou et al. conducted an analysis of Kerr–Melvin black holes illuminated by thin accretion disks, demonstrating that their inner shadow morphology and critical curve structure can serve as diagnostic tools for inferring ambient magnetic fields [14]. Concurrently, Guo et al. examined the optical properties of Kerr–Newman black holes, investigating the effects of spin, charge, and observer inclination on the appearance of black hole images [28]. Analytical studies of photon rings in charged spacetimes [29,30] suggest that even small charges ($Q/M \sim 0.1$) can induce observable asymmetries in the shadow structure. These predictions are supported by comprehensive ray-tracing frameworks [16].

Given the structural similarities between the Kerr–Newman metric in general relativity and other related solutions, it is a natural progression to extend these analyses to Kerr–Sen black holes. Within the framework of the Einstein–Maxwell–Dilaton–Axion (EMDA) model, the Kerr–Sen metric describes a stationary and axisymmetric black hole with geometric features that deviate markedly from those of Kerr–Newman black holes, primarily due to the inclusion of additional scalar and pseudoscalar fields such as the dilaton and axion [31,32]. Investigating the spacetime structure and potential observational imprints of Kerr–Sen black holes is of particular relevance, as they provide a potential avenue for probing the low-energy phenomenology of string theory in the strong-field regime [33,34].

Although numerous theoretical investigations have been conducted, the optical appearance of Kerr–Sen black holes,

especially those surrounded by thin accretion disks, remains insufficiently characterized in the literature [19]. Analyzing their inner shadows and photon rings may provide new diagnostic tools for constraining external electromagnetic fields and quantifying the influence of electric charge on observational signatures [35].

The Kerr–Sen black hole, an extension of the Kerr black hole within the framework of string theory, has received limited attention regarding its observational characteristics [36,37]. This study aims to address this gap by investigating photon trajectories in the vicinity of a Kerr–Sen black hole surrounded by a thin accretion disk. Employing general relativity and ray-tracing techniques, we examine the potential observational signatures of this theoretically motivated black hole solution [38]. We compute photon trajectories and analyze the effects of gravitational lensing and particle orbits, as well as the frequency-dependent variations in redshift and intensity of the black hole at 230 GHz and 86 GHz [39]. This research not only advances our understanding of Kerr–Sen black holes but also provides new avenues for testing general relativity and exploring high-energy astrophysical phenomena, thereby offering novel perspectives for future observational campaigns.

The structure of this paper is as follows: in Sect. 2, we review the Kerr–Sen black hole and utilize a semi-analytic approach to compute the photon deflection and the lensing ring in its vicinity. Section 3 focuses on deriving the redshift distribution associated with the Kerr–Sen black hole and presents simulations at various observational frequencies, accounting for both prograde and retrograde accretion disk configurations. Finally, we summarize the key findings of this study.

## 2 Ray-tracing of the kerr–Sen black hole

In this section, we review the dynamics of the Kerr–Sen black hole and explore the ray-tracing within its context. The metric of the Kerr–Sen black hole, expressed in Boyer–Lindquist coordinates, is given by [31,36,37]:

$$ds^2 = -\frac{1-2Mr}{\Sigma}dt^2 + \frac{\Sigma}{\Delta}dr^2 + \Sigma d\theta^2 - \frac{4aMr}{\Sigma}\sin^2\theta dt d\phi$$
$$+ \sin^2\theta\left(r(r+r_0)+a^2+\frac{2Mra^2\sin^2\theta}{\Sigma}\right)d\phi^2, (1)$$

in which

$$\Sigma = r(r+r_0)+a^2\cos^2\theta, \quad \Delta = r(r+r_0)+a^2-2Mr. \tag{2}$$

In the equations above, $M$ denotes the mass of the black hole, and $a$ represents the spin parameter, corresponding to the angular momentum per unit mass [33]. The expansion parameter $r_0$ is determined by the electric charge $Q$ and the spin parameter $a$, with $a = \frac{J}{M}$ and $r_0 = \frac{Q^2}{M}$ [40]. As can be seen from Eq. (1), setting the spin parameter $a$ to zero yields a static, spherically symmetric dilaton black hole solution [36]. Moreover, in the limiting case where $r_0 \to 0$, corresponding to vanishing electric charge, the Kerr–Sen solution smoothly transitions to the Kerr black hole [31]. The location of the event horizon $r_h$ is determined by the condition $\Delta = 0$, and for the Kerr–Sen black hole, it is given by [33,40]:

$$r_h = M - \frac{r_0}{2} \pm \sqrt{\left(M - \frac{r_0}{2}\right)^2 - a^2}. \tag{3}$$

Equation (3) defines the theoretical bounds for the inner and outer event horizons of the black hole, expressed as [33,40]:

$$0 \leq \frac{r_0}{M} \leq 2\left(1 - \frac{a}{M}\right) \quad \text{or}$$
$$-\left(1 - \frac{r_0}{2M}\right) \leq \frac{a}{M} \leq 1 - \frac{r_0}{2M},$$

the spin parameter $a$ is subject to the condition that it must not exceed the black hole mass $M$, i.e. $a \leq M$. From this constraint, we can derive the allowed range for the expansion parameter $r_0$, given by [31,33,36,41]:

$$0 \leq \frac{r_0}{M} \leq 2.$$

### 2.1 Photon trajectory near a Kerr–Sen black hole

To investigate the trajectory of photons around a Kerr–Sen black hole, it is essential to analyze the dynamics of particles within the corresponding spacetime. In particular, the behavior of massless particles such as photons provides key insights into the geometry and causal structure of the black hole environment. In this context, the motion of photons is governed by the geodesic equation, which describes their trajectories under the influence of spacetime curvature. Specifically, the geodesic equation takes the form:

$$\frac{d^2 x^\mu}{d\lambda^2} + \Gamma^\mu_{\alpha\beta} \frac{dx^\alpha}{d\lambda} \frac{dx^\beta}{d\lambda} = 0, \tag{4}$$

where $\lambda$ denotes the affine parameter, and $\Gamma^\mu_{\alpha\beta}$ are the Christoffel symbols, which encapsulate the geometric properties of spacetime. Although, in principle, the geodesic equation can be solved analytically or numerically given appropriate initial conditions and a specified metric, this approach is often computationally intensive. As a more tractable alternative, geodesic motion can be studied more efficiently using the Hamilton–Jacobi formalism, which simplifies the analysis by reducing the second-order differential equations to a single first-order partial differential equation. In the spacetime of the Kerr–Sen black hole, the Hamilton–Jacobi equation for a test particle takes the following form

[42–45]:

$$\frac{\partial S}{\partial \lambda} = -\frac{1}{2} g^{\mu\nu} \frac{\partial S}{\partial x^\mu} \frac{\partial S}{\partial x^\nu}. \tag{5}$$

Analogous to the analysis of spherically symmetric black holes, the trajectory of photons around a rotating black hole can be characterized by conserved quantities. In this case, the motion of photons is governed by three constants of motion: $p_t = -E$, $p_\phi = L_z$, and the Carter constant,

$$C = p_\theta^2 - \cos^2\theta \left(a^2 p_t^2 - p_\phi^2 \csc^2\theta\right),$$

where $E$, $L_z$, and $C$ correspond to the energy, the axial component of the angular momentum, and the Carter constant, respectively. The presence of the Carter constant reflects an additional conserved quantity arising from a hidden symmetry of the spacetime that is not associated with a Killing vector field. It plays a crucial role in governing the motion in the polar direction, i.e., the evolution of the $\theta$-coordinate, and ensures that the geodesic equations remain separable in the Kerr–Sen geometry.

$$\xi = \frac{L}{E}, \quad \eta = \frac{C}{E^2}. \tag{6}$$

For the Kerr–Sen black hole, the four-momentum $p^\mu$ of a photon moving along its orbit is described by the following equations [30,31,36,46]:

$$\frac{\Sigma}{E} p^r = \pm\sqrt{R(r)}, \tag{7}$$

$$\frac{\Sigma}{E} p^\theta = \pm\sqrt{\Theta(r)}, \tag{8}$$

$$\frac{\Sigma}{E} p^t = \frac{(r(r + r_0) + a^2)^2 - \Delta a^2 \sin^2\theta - 2Mra\xi}{\Delta}, \tag{9}$$

$$\frac{\Sigma}{E} p^\phi = \frac{(\Sigma - 2Mr)\xi + 2Mra}{\Delta \sin^2\theta}. \tag{10}$$

In these equations, $R(r)$ and $\Theta(r)$ represent the radial and angular potentials, respectively [32,43–45]:

$$R(r) = a^2 \xi^2 + (r(r + r_0) + a^2)^2$$
$$\quad - 4Mra\xi - \Delta(\eta + a^2 + \xi^2), \tag{11}$$

$$\Theta(\theta) = \eta + a^2 \cos^2\theta - \xi^2 \cot^2\theta. \tag{12}$$

The $\pm_r$ and $\pm_\theta$ denote the components $p^r$ and $p^\theta$, respectively, where $r$ and $\theta$ are the coordinates at the inflection points of the photon trajectory. These inflection points correspond to turning points in the radial and polar motion of the photon, which occur at the zeros of the radial potential $R(r)$ and the angular potential $\Theta(\theta)$, respectively [43,44]. To describe photon propagation from a localized emitter to a distant observer, consider a photon emitted from a source at coordinates $(t_s, r_s, \theta_s, \phi_s)$ and received by an observer at coordinates $(t_o, \infty, \theta_o, \phi_o)$. To simplify the analysis of null geodesics, we introduce the Mino time $\tau$ to parameterize the trajectory. The geodesic equations in terms of Mino time $\tau$ can then be written as follows [47–49]:

$$\frac{dx^\mu}{d\tau} \equiv \frac{\Sigma}{E} p^\mu. \tag{13}$$

The integral forms of the equations of motion can be further derived from Eqs. (7)–(12) [47]:

$$\Delta t = t_o - t_s = I_t + a^2 G_t, \tag{14}$$

$$\tau = I_r = G_\theta, \tag{15}$$

$$\Delta\phi = \phi_o - \phi_s = I_\phi + \xi G_\phi, \tag{16}$$

where

$$I_t = -\int_{r_s}^{r_o} \frac{(2Mr - rr_0)(a^2 + r^2 - a\xi) + \Delta(r)r^2}{\pm_r \Delta(r)\sqrt{R(r)}} dr, \tag{17}$$

$$I_r = -\int_{r_s}^{r_o} \frac{dr}{\pm_r \sqrt{R(r)}}, \tag{18}$$

$$I_\phi = -\int_{r_s}^{r_o} \frac{a(2Mr - r_0 - a\xi)}{\pm_r \Delta(r)\sqrt{R(r)}} dr, \tag{19}$$

$$G_t = -\int_{\theta_s}^{\theta_o} \frac{\cos^2\theta}{\pm_\theta \sqrt{\Theta(\theta)}} d\theta, \tag{20}$$

$$G_\theta = -\int_{\theta_s}^{\theta_o} \frac{d\theta}{\pm_\theta \sqrt{\Theta(\theta)}}, \tag{21}$$

$$G_\phi = -\int_{\theta_s}^{\theta_o} \frac{\csc^2\theta}{\pm_\theta \sqrt{\Theta(\theta)}} d\theta. \tag{22}$$

The integrals $I_t$, $I_r$, and $I_\phi$ represent the path integrals between the source and observer coordinates, $x_s^\mu$ and $x_o^\mu$, respectively. These integrals correspond to the coordinate differences accumulated along the null geodesic in the temporal, radial, and azimuthal directions. For the angular integral, where $\eta > 0$, the geodesic can be characterized by the inflection points $\theta_\pm$ in the polar motion, which occur symmetrically with respect to the equatorial plane, as shown in [31,36,50]:

$$\theta_\pm = \arccos(\mp\sqrt{\omega_\pm}), \tag{23}$$

and

$$\omega_\pm = \frac{1}{2} - \frac{\eta + \xi^2}{2a^2} \pm \sqrt{\frac{\eta}{a^2} + \frac{1}{4}\left(1 - \frac{\eta - \xi^2}{a^2}\right)^2}. \tag{24}$$

To avoid the singularity of spherical coordinates, we assume $0 < \theta < \pi$, and define the angular potential in terms of $\omega = \cos^2\theta$. It follows that $\arccos(\mp\sqrt{\omega_\pm})$ are the four roots of $\Theta(\theta)$. The special cases occur if and only if $\omega_+ = 0$, $\omega_- = 0$, or $\omega_+ = \omega_-$. These conditions divide the $(\xi, \eta)$-plane into several distinct regions. In each such region, the "characteristics" of the potential-namely, the number of real roots and the sign of the potential on either side of those roots-remain invariant. Therefore, the characteristics of each region can be determined by evaluating a single representative point within it.

In the above formulation, the positive angular integral is expressed in terms of elliptic integrals, with the number of turning points encountered along the trajectory being calculated. We obtain [47,50]:

$$G_\theta = \frac{1}{a\sqrt{-\omega_-}}[2zK \pm_s F_s \mp_o F_o], \tag{25}$$

$$G_\phi = \frac{1}{a\sqrt{-\omega_-}}[2z\Pi \pm_s \Pi_s \mp_o \Pi_o], \tag{26}$$

$$G_t = -\frac{2\omega_-}{a\sqrt{-\omega_-}}[2zE' \pm_s E'_s \mp_o E'_o], \tag{27}$$

where $K$, $\Pi$, and $E'$ represent the complete elliptic integrals of the first, third, and second kinds, respectively, while $K_i$, $\Pi_i$, and $E'_i$ are the corresponding incomplete elliptic integrals. These integrals arise naturally in the evaluation of the geodesic equations, particularly when separating the motion into radial and angular components. For the Kerr–Sen black hole, the radial potential depends on $r_0$, which is related to the black hole's charge $Q$. Specifically, $r_0 = Q^2/M$ introduces a charge-dependent deformation to the potential, modifying the structure of photon orbits compared to the Kerr case. To determine the two specific parameters that characterize the ray (e.g., impact parameters $\xi$ and $\eta$), we must solve Eqs. (2) and (11) to find the roots of the radial potential, i.e.,

$$R(r) = r^4 + Ar^3 + Br^2 + Dr + U, \tag{28}$$

with

$$A = 2r_0, \quad B = r_0^2 - \eta + a^2 - \xi^2, \tag{29}$$

$$D = 2a^2 r_0 - \eta r_0 - a^2 r_0 - \xi^2 r_0 - 4Ma\xi, \tag{30}$$

$$U = a^2\xi^2 + a^4.$$

The four roots of the equation (28) are determined by:

$$r_1 = \frac{1}{2}j - \frac{1}{2}\sqrt{-\frac{2B}{3} - j^2 - \frac{2D}{j}}, \tag{31}$$

$$r_2 = \frac{1}{2}j + \frac{1}{2}\sqrt{-\frac{2B}{3} - j^2 - \frac{2D}{j}}, \tag{32}$$

$$r_3 = -\frac{1}{2}j - \frac{1}{2}\sqrt{-\frac{2B}{3} - j^2 + \frac{2D}{j}}, \tag{33}$$

$$r_4 = -\frac{1}{2}j + \frac{1}{2}\sqrt{-\frac{2B}{3} - j^2 + \frac{2D}{j}}, \tag{34}$$

where the parameter $j$ is given by:

$$j = \sqrt{\frac{-\frac{B}{3} + \tilde{\omega}_+ + \tilde{\omega}_-}{2}}. \tag{35}$$

Here, $\tilde{\omega}_\pm$ are defined as in Eq. (24). Finally, for an observer at infinity, the photon will encounter an inflection point at $r_4$ along its trajectory. For rays without turning points, the radial integrals $I_r$, $I_\phi$, and $I_t$ are single-valued functions of $r_s$. However, for rays with turning points, these radial integrals must be treated as double-valued in order to account for whether the ray has reached a turning point. We denote the number of turning points experienced by a photon (a segment of the ray) by $\omega \in \{0, 1\}$. The expression for the radial integral is further written as:

$$I_i \sim -\int_{r_s}^{r_o} dr \sim \int_{r_s}^{r_o} \cdot + 2\omega \int_{r_t}^{r_s} dr \cdots \qquad (36)$$

In this way, we are able to derive the explicit expressions for the radial integrals:

$$I_r = \int_{r_s}^{\infty} \frac{dr}{\sqrt{R(r)}} + 2\omega \int_{r_4}^{r_s} \frac{dr}{\sqrt{R(r)}} \qquad (37)$$

For the two distinct light sources, we consider the following scenarios:

$$I_r^{total} = 2\int_{r_s}^{\infty} \frac{dr}{\sqrt{R(r)}}, \quad r_+ < r_4 \in \mathbb{R}, \qquad (38)$$

$$I_r^{total} = \int_{r_+}^{\infty} \frac{dr}{\sqrt{R(r)}}, \quad \text{otherwise.} \qquad (39)$$

In the above equation, the parameter $r_4$ must be real-valued and lie outside the event horizon to correspond to a physically meaningful turning point. Following the analytical method proposed by Gralla et al., the inverse formulation of the path integral can be obtained by exchanging the spatial positions of the observer and the photon source [50]. This exchange effectively reverses the direction of integration along the photon's path, resulting in a sign change of $I_r$ to its negative counterpart (i.e., $-I_r$). This yields an expression for the photon source radius $r_s$, namely:

$$r_s = \frac{r_4 r_{31} - r_3 r_{41} \, sn^2\left(\frac{1}{2}\sqrt{r_{31}r_{42}} \, I_r\right) - \mathcal{F}_o \left|_{r_{31}r_{42}}^{r_{32}r_{41}}\right.}{r_{31} - r_{41} \, sn^2\left(\frac{1}{2}\sqrt{r_{31}r_{42}} \, I_r\right) - \mathcal{F}_o \left|_{r_{31}r_{42}}^{r_{32}r_{41}}\right.}, \qquad (40)$$

where the function $\mathcal{F}_o$ is explicitly defined as [50]

$$\mathcal{F}_o = F\left(\arcsin\sqrt{\frac{r_{31}}{r_{41}}} \; \left|\; \frac{r_{32}r_{41}}{r_{31}r_{42}}\right.\right). \qquad (41)$$

Analogous to the treatment of spherically symmetric black holes, elliptic integrals can be effectively employed to derive the geodesic equations governing particle motion in the Kerr–Sen spacetime. These integrals facilitate the exact analytical treatment of the equations of motion, particularly when the geodesics involve non-trivial turning points in both radial and angular directions. Given the possibility of particles executing multiple revolutions near the Kerr–Sen black hole, accurately calculating the orbital winding number becomes crucial. This quantity characterizes the number of azimuthal cycles a particle undergoes as it travels from the source to the observer, and plays a key role in the analysis of strong gravitational lensing and photon ring structures. We can obtain:

$$n = \frac{G_\theta}{2\int_{\theta_-}^{\theta_+} \sqrt{\Theta(\theta)} \, d\theta} = \frac{a\sqrt{-\omega_-}}{4K} I_r. \qquad (42)$$

By combining Eqs. (25) and (42), we obtain an explicit relation that connects the orbital fraction $n$ with the number of turning points $z$ for particle trajectories around a Kerr–Sen black hole.

$$n = \frac{z}{2} \pm_o \frac{1}{4}\left[(-1)^z \frac{F_s}{K} \frac{F_o}{K}\right]. \qquad (43)$$

In the general case characterized by two independent parameters, the radial Eq. (11) admits four distinct roots, with the real-valued subset corresponding precisely to the radial turning points. Analogous to the Schwarzschild black hole case, the critical impact parameter defines the location of the photon ring. For rotating black holes, when the impact parameter satisfies the criticality conditions, the corresponding turning points ($\tilde{r}$) are determined by solving the system $R(\tilde{r}) = 0$ and $R'(\tilde{r}) = 0$. Based on these conditions, two distinct sets of solutions arise, as derived in Ref. [46,48]:

$$\tilde{\eta} = -\frac{\tilde{r}^2(\tilde{r} + r_0)}{a^2},$$

$$\tilde{\xi} = \frac{r_4 r_{31} - r_3 r_{41} \, sn^2\left(\frac{1}{2}\sqrt{r_{31}r_{42}} \, I_r\right) - \mathcal{F}_o \left|_{r_{31}r_{42}}^{r_{32}r_{41}}\right.}{a}. \qquad (44)$$

For the Kerr–Sen black hole, the specific expressions for the two scenarios that can be derived are as follows:

1.

$$\tilde{\eta} = -\frac{\tilde{r}^2(\tilde{r} + r_0)^2}{a^2}, \quad \tilde{\xi} = \frac{a^2 + (\tilde{r} + r_0)\tilde{r}}{a}. \qquad (45)$$

2.

$$\tilde{\eta} = \frac{-\tilde{r}^2\left[-8a^2M(2\tilde{r} + r_0) + ((\tilde{r} + r_0)(2\tilde{r} + r_0) - 2M(3\tilde{r} + r_0))^2\right]}{a^2}, \qquad (46)$$

$$\tilde{\xi} = \frac{a^2(2M + \tilde{r} + r_0) + \tilde{r}(\tilde{r} + r_0)(2\tilde{r} + r_0) - 2M(3\tilde{r} + r_0)}{a(2M - \tilde{r} - r_0)}. \qquad (47)$$

When a geodesic trajectory intersects the equatorial plane, the squared momentum component normal to this plane must be strictly nonnegative, thus requiring the Carter constant $C$ (or equivalently, the parameter $\tilde{\eta}$) to satisfy $\tilde{\eta} > 0$. This condition ensures that the motion possesses a real angular component in the $\theta$-direction and excludes trajectories that are confined strictly to the equatorial plane without polar variation. Consequently, this constraint eliminates scenarios characterized by negative values of $\tilde{\eta}$, leaving the first class of geodesics perpendicular to the equatorial plane and leaving only the second class as physically relevant. In this context, the second class refers to generic photon trajectories with nonzero inclination angles. Specifically, for the Kerr–Sen black hole, setting $Q = 0$ yields a simplified analytical expression in terms of a critical cubic equation whose solution explicitly provides the radial turning points [36,47]:

$$\tilde{r}_\pm = M \left[ 1 + \cos \left( \frac{2}{3} \arccos \left( \pm \frac{a}{M} \right) \right) \right]. \tag{48}$$

We now revisit Eq. (46). In the black hole spacetime, the parameter $\tilde{\eta}$ emerges as a quartic polynomial function of the radius $\tilde{r}$, which renders analytical approaches relying solely on trigonometric simplifications insufficiently effective. This complexity arises from the nontrivial dependence of the Carter constant on both the radial potential and the spin-induced geometry of the Kerr–Sen spacetime. Therefore, numerical methods must be employed to precisely solve this equation, ensuring the accurate determination of the Kerr–Sen black hole's critical curve. Moreover, as previously established, the condition $\tilde{\eta} > 0$ is required for geodesics intersecting the equatorial plane. This physical constraint restricts the domain of admissible solutions and guarantees real-valued turning points in the angular sector. Under this constraint, Eq. (46) simplifies explicitly into the following quartic polynomial equation:

$$-\tilde{r}^4 + (6M - 3r_0)\tilde{r}^3 + \left( 11Mr_0 - 9M^2 - \frac{13}{4}r_0^2 \right) \tilde{r}^2$$
$$+ \left( 4a^2 M + 6Mr_0^2 - 6M^2 r_0 - \frac{3}{2}r_0^3 \right) \tilde{r}$$
$$- \left( \frac{1}{4}r_0^4 - Mr_0^3 + M^2 r_0^2 - 2a^2 Mr_0 \right) = 0. \tag{49}$$

The radial potential $R(r)$ can be expressed as a quartic polynomial, and thus written as $R(r) = (r - \tilde{r}_1)(r - \tilde{r}_2)(r - \tilde{r}_3)(r - \tilde{r}_4)$, where the roots satisfy $\tilde{r}_1 < \tilde{r}_2 < \tilde{r}_3 < \tilde{r}_4$, and obey the constraint $\tilde{r}_1 + \tilde{r}_2 + \tilde{r}_3 + \tilde{r}_4 = 0$. The detailed derivation is presented in Appendix B of the paper [51].

Figure 1 illustrates the functional relationship between $\tilde{r}$ and $\tilde{\eta}$, showing four distinct roots that satisfy the ordering $\tilde{r}_1 < \tilde{r}_2 < \tilde{r}_h < \tilde{r}_3 < \tilde{r}_4$. Given the event horizon located at radius $\tilde{r}_h$, this ordering naturally dictates the identification of the physically relevant roots as $\tilde{r}_3$ and $\tilde{r}_4$, which are denoted as $\tilde{r}_-$ and $\tilde{r}_+$, respectively.

## 2.2 Lensing bands and photon orbits in Kerr–Sen spacetime

Consider an observer located at spatial infinity. In the observer's image plane, the positions of photon impacts on the screen can be described by the orthogonal coordinates $(\alpha, \beta)$. The observer's sky is projected onto a two-dimensional plane, whose Cartesian coordinates are proportional to the sines and cosines of the photon's incident angles. In the context of the Kerr–Sen black hole, where particle

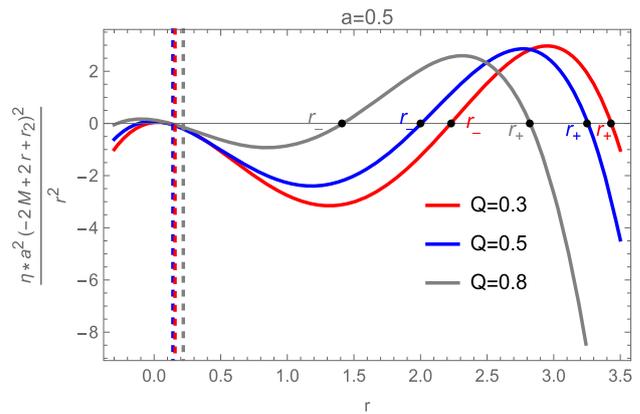

**Fig. 1** The dependence of $\tilde{\eta}$ on the radial coordinate $\tilde{r}$. The dashed line marks the location of the black hole's event horizon at $r_h$, while the solid curves illustrate the functional behavior of $\tilde{\eta}$ as a function of $\tilde{r}$. For different values of the charge $Q$, the curves correspond to parameter choices of $a = 0.5$ and $M = 1$, with the red, blue, and gray lines representing $Q = 0.3$, $Q = 0.5$, and $Q = 0.8$, respectively

trajectories conserve two specific constants of motion, the parameters $\alpha$ and $\beta$ are expressed as follows:

$$\alpha = -\frac{\xi}{\sin \theta_0}, \tag{50}$$

$$\beta = \pm_o \sqrt{\Theta(\theta)} = \pm_o \sqrt{\eta + a^2 \cos^2 \theta_0 - \xi^2 \cot^2 \theta_0}. \tag{51}$$

Note that in the aforementioned equations, the parameter $\theta_0$ represents the inclination angle of the observer. Using these equations, the photon lensing ring-composed of photons with critical impact parameters $\tilde{\xi}(\tilde{r})$ and $\tilde{\eta}(\tilde{r})$-can be explicitly obtained. This lensing ring is naturally represented by the coordinates $\tilde{\alpha}(\tilde{r})$ and $\tilde{\beta}(\tilde{r})$ on the observer's sky. From Eqs. (46), (47), (50) and (51), we obtain Fig. 2.

Figure 2 illustrates the photon lensing rings of the Kerr–Sen black hole at various observational inclination angles, with an image resolution of $1024 \times 1024$ pixels. The solid white line denotes the critical photon curve, while the first through fourth columns correspond to four distinct inclination angles: $\theta_0 = 17°$, $53°$, $75°$, and $150°$. Notably, the angle $\theta_0 = 17°$ closely approximates the observational inclination angle employed for M87* by the EHT. As the inclination angle $\theta_0$ increases from small values towards $\theta_0 < \pi/2$, the critical curve progressively elongates in the southwest direction. Upon further increase, surpassing the equatorial plane and approaching inversion at larger angles, the orientation of the critical curve undergoes an almost complete reversal.

Consider the particle orbits around the Kerr–Sen black hole, where our discussion primarily focuses on sources located on the equatorial plane. In this scenario, the observer is positioned at a specific observational inclination angle $\theta_0 \neq 0$, which results in the first kind of incomplete elliptic

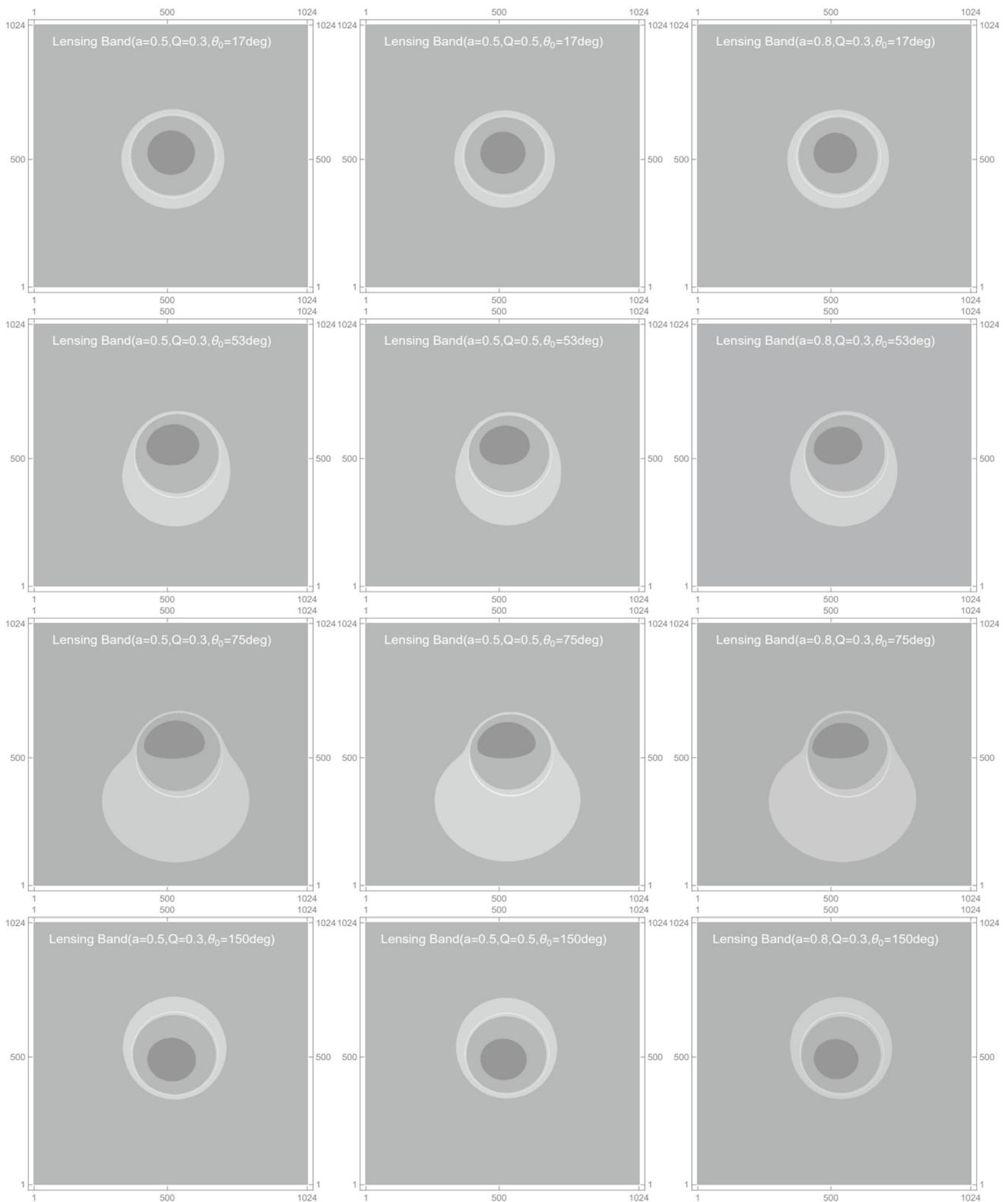

**Fig. 2** Photon lensing bands surrounding the Kerr–Sen black hole are shown for different spin parameters $a$, electric charge parameters $Q$, and observational inclination angles $\theta_0$. Each column, from left to right, corresponds to the parameter sets: ($a = 0.5$, $Q = 0.3$), ($a = 0.5$, $Q = 0.5$), and ($a = 0.8$, $Q = 0.3$). Each row, from top to bottom, corresponds to observational inclination angles: 17°, 53°, 75°, and 150°. The dark gray and light gray shaded regions correspond to photon trajectories intersecting the black hole's photon sphere once and twice, respectively. The solid white line denotes the critical photon curve, and the innermost region bounded by this curve represents the "inner shadow", a distinctive feature of the Kerr–Sen black hole. All images assume a normalized black hole mass of $M = 1$

integral $F_s$ becoming zero. We can then derive:

$$\sqrt{-w_-a^2}I_r + \text{sign}(\beta)F_o = 2zK, \quad (52)$$

where $F_o$ represents the incomplete elliptic integral of the first kind, and $K$ denotes the complete elliptic integral of the first kind. The above equation provides the relationship between the different radial positions $r_s$ and the coordinates $\alpha$ and $\beta$.

## 3 Image of the Kerr–Sen black hole

Next, we focus on the image of a Kerr–Sen black hole surrounded by a geometrically thin and optically thin accretion disk. In this context, we consider a standard relativistic disk model, where the disk material follows nearly circular geodesics in the equatorial plane. We assume that the accretion disk lies entirely within the equatorial plane, with its innermost region defined by the innermost stable circular orbit (ISCO). Physically, the ISCO delineates the boundary between stable and unstable circular orbits; particles located within this radius are subject to radial perturbations, inevitably migrating inward and ultimately plunging into the black hole. The location of the ISCO is sensitive to the black hole's spin and charge parameters, and thus plays a crucial role in shaping the observed image of the accretion flow.

Considering the steep increase in orbital velocities of test particles as they approach the ISCO near a Kerr–Sen black hole, the classical accretion disk model can be further refined. This modification effectively shifts the inner boundary of the accretion disk closer to the event horizon of the black hole. In other words, this refined accretion disk model positions its innermost boundary nearer to the event horizon while keeping its outer boundary beyond the ISCO. This refinement introduces two notable considerations:

1. A comprehensive reevaluation of gravitational redshift and radiative transfer processes becomes necessary near the ISCO region.
2. Implementation of ray-tracing techniques requires integration over a significantly expanded radial region.

Moreover, we investigate the image of a Kerr–Sen black hole encompassed by both prograde and retrograde accretion disks. It is important to note that the ISCO demarcates stable circular orbits from unstable plunging trajectories. Particles located outside the ISCO can maintain stable circular orbits, whereas particles within this radius inevitably spiral inward toward the black hole. The radial equation governing particle

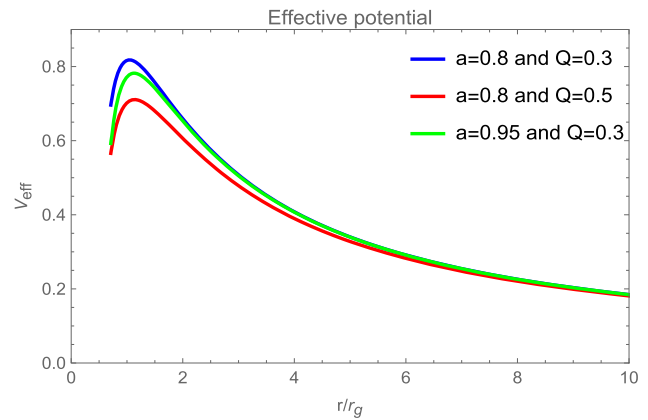

**Fig. 3** Radial profiles of the effective potential for test particles orbiting the Kerr–Sen black hole. The blue, red, and green curves correspond to parameter sets $(a = 0.8, Q = 0.3)$, $(a = 0.8, Q = 0.5)$, and $(a = 0.95, Q = 0.3)$, respectively. The black hole mass is normalized to $M = 1$

motion within the equatorial plane is explicitly given by [14]:

$$u^r = -\sqrt{-\frac{V(r, E, L)}{g^{rr}}}, \quad (53)$$

with the effective potential given by:

$$V(r, E, L) = \left( E^2 g^{tt} + 2EL g^{t\phi} + L^2 g^{\phi\phi} + 1 \right)\Big|_{\theta=\frac{\pi}{2}}. \quad (54)$$

Revisiting Eq. (54), we examine particle trajectories around the Kerr–Sen black hole via the effective potential. Figure 3 illustrates the radial dependence of the effective potential for various spin parameters $a$ and electric charges $Q$. It is manifest from the curves that both parameters exert a significant influence on the structure of the effective potential. Specifically, increasing either the spin parameter $a$ or the electric charge $Q$ shifts the location of the potential peak outward while simultaneously enhancing its magnitude, thereby effectively modifying the region where particle orbits remain stable. These results demonstrate that variations in spin and electric charge parameters fundamentally reshape the underlying spacetime geometry, consequently altering the gravitational environment experienced by test particles orbiting the Kerr–Sen black hole.

By applying the critical conditions of the effective potential, we determine the radius of the ISCO for particles orbiting the Kerr–Sen black hole. Specifically, the ISCO radius $r_{\text{ISCO}}$ is defined by the simultaneous conditions $V(r_{\text{ISCO}}, E, L) = 0$ and $V'(r_{\text{ISCO}}, E, L) = 0$. For radii $r > r_{\text{ISCO}}$, particle motion is governed by stable circular orbit conditions imposed by the effective potential. However, for radii within the ISCO, i.e., $r < r_{\text{ISCO}}$, the radial motion of plunging par-ticles is explicitly governed by the following equation [48]:

$$u^r = -\sqrt{-\frac{V(r, E_{ISCO}, L_{ISCO})}{g^{rr}}}. \quad (55)$$

### 3.1 Zero-angular–momentum observer

To model a fisheye camera and describe the observer's local reference frame, we select a standard orthonormal tetrad, providing a local reference frame at each point in spacetime. For an observer situated within the Kerr–Sen black hole spacetime, this orthonormal tetrad can be expressed as:

$$e_t = \delta \partial_t + \chi \partial_\phi, \quad e_r = \frac{1}{\sqrt{g_{rr}}} \partial_r, \quad (56)$$

$$e_\theta = \frac{1}{\sqrt{g_{\theta\theta}}} \partial_\theta, \quad e_\phi = \frac{1}{\sqrt{g_{\phi\phi}}} \partial_\phi, \quad (57)$$

where

$$\delta = \sqrt{\frac{g_{\phi\phi}}{g_{t\phi}^2 - g_{tt}g_{\phi\phi}}}, \quad \chi = -\frac{g_{t\phi}}{g_{\phi\phi}} \sqrt{\frac{g_{\phi\phi}}{g_{t\phi}^2 - g_{tt}g_{\phi\phi}}}. \quad (58)$$

In the frame of an observer with zero angular momentum, the trajectories of photons are reversible, implying that any photon arriving at the observer's position can be completely captured. For such an observer, the photon's four-momentum can be measured using the observer's local tetrad. The photon's four-momentum $p^\mu$ is expressed as:

$$p^\mu = \eta^{\mu\nu} e_\nu^\xi k_\xi. \quad (59)$$

The complete four-momentum is given by [25,37,50]:

$$P^t = E(\delta - \chi\xi), \quad P^r = E \frac{1}{\sqrt{g_{rr}}} \frac{\pm_r \sqrt{R(r)}}{\Delta_r}, \quad (60)$$

$$P^\theta = E \frac{1}{\sqrt{g_{\theta\theta}}} \frac{\pm_\theta \sqrt{\Theta(\theta)}}{\Delta_\theta}, \quad P^\phi = E \frac{\xi}{\sqrt{g_{\phi\phi}}}. \quad (61)$$

Figures 4 and 5 display images of the inner shadow of the Kerr–Sen black hole, generated using the fisheye camera principle. The photon trajectories are distinctly divided into four quadrants in the figures, exhibiting varying degrees of clarity. As the charge and spin parameters increase, a pronounced bending of light rays is observed near the Kerr–Sen black hole. Notably, the distortion induced by increasing the spin parameter is substantially more significant than that caused by an increase in charge.

To better illustrate the variations in the inner shadow of the Kerr–Sen black hole, we have magnified the inner shadow regions from Figs. 4 and 5, presenting them in greater detail in Figs. 6 and 7. The results demonstrate that as both the charge and spin parameters increase, the inner shadow undergoes progressively more pronounced distortions. In particular, variations in the spin parameter produce a significantly stronger impact on the shape and deformation of the inner shadow compared to changes in the charge parameter.

In Figs. 8 and 9, we present the inner shadow of the Kerr–Sen black hole under varying observation inclinations, charge values, and spin parameters. By superimposing three inner shadow images (with controlled variables) for comparison, we observe the following trends:

- As the observation inclination increases from 0 to 90 degrees, the inner shadow gradually becomes flattened; beyond 90 degrees, the deformation progressively diminishes.
- As the charge increases, the weakening effect of the gravitational field induced by the charge intensifies, leading to a reduction in the size of the inner shadow.
- A similar reduction in the size of the inner shadow occurs as the spin parameter grows, driven by the enhanced frame-dragging effect of spacetime caused by the black hole's rotation.
- Critically, across all observation inclinations, the influence of the spin parameter on the deformation of the inner shadow remains substantially stronger than that of the charge.

### 3.2 Intensity and redshift

When considering the light emitted from the accretion disk that reaches the observer's plane, careful attention must be paid to changes in luminosity caused by various factors such as scattering, absorption, the Doppler effect, and gravitational redshift. In this analysis, we neglect the refractive effects arising from the accretion disk medium, assuming it to be optically thin and non-dispersive. The observed luminosity can be expressed as:

$$\frac{d}{d\lambda}\left(\frac{I_\nu}{\nu^3}\right) = \frac{J_\nu - \kappa_\nu I_\nu}{\nu^2}. \quad (62)$$

Here, $\lambda$ is the affine parameter, and $I_\nu$, $J_\nu$, and $\kappa_\nu$ represent the intensity, emissivity, and absorption coefficient at frequency $\nu$, respectively. Since light propagates in a vacuum, both $J_\nu$ and $\kappa_\nu$ are zero. Therefore, the value of $\frac{I_\nu}{\nu^3}$ remains constant along the geodesic path.

From the above discussion, we know that the accretion disk is thin and transparent both geometrically and optically, and we can integrate the photon trajectory to determine the intensity at each position on the observer's screen. Therefore, we obtain:

$$I_{\nu_o} = \sum_{m=1}^{N_{max}} \left(\frac{\nu_o}{\nu_m}\right)^3 \frac{J_m}{\tau_{m-1}}\left[\frac{1 - e^{-\kappa_m f_m}}{\kappa_m}\right]. \quad (63)$$

Here, $\nu_o$ represents the frequency on the observer's screen, $\nu_m$ denotes the frequency observed in the local rest frame of the accretion disk, and $\tau_k$ represents the optical depth of the photons emitted from point $k$. With the characteristics of the accretion disk, we can simplify Eq. (63) as:

$$I_{v_o} = \sum_{m=1}^{N_{max}} f_m g^3(r_m) J_{model}(r_m).$$ (64)

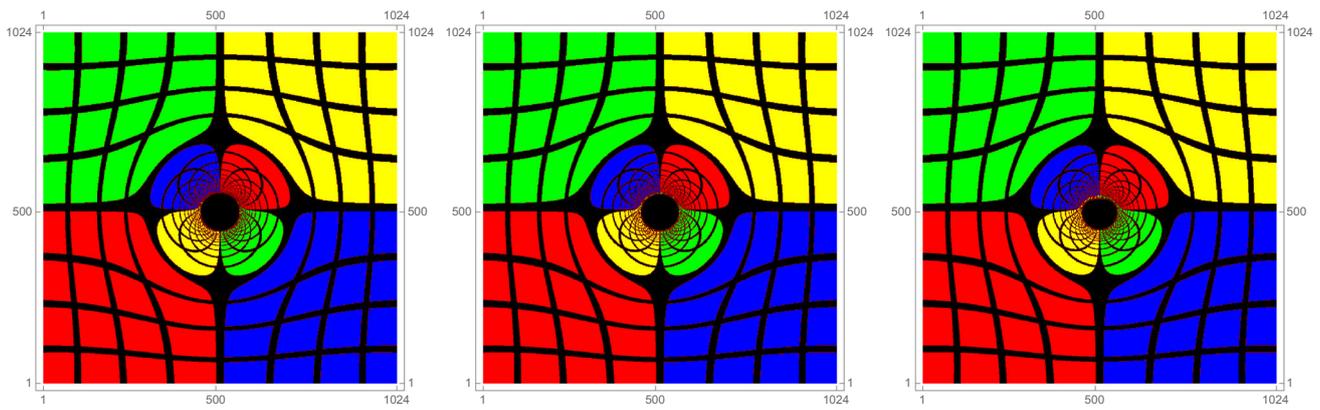

**Fig. 4** The shadow of the Kerr–Sen black hole is analyzed using a numerical ray-tracing method. The figure above illustrates the results obtained through this method for different parameter configurations of the Kerr–Sen black hole. From left to right, the selected parameter sets are $a = 0.5$, $Q = 0.3$, $a = 0.5$, $Q = 0.5$, and $a = 0.5$, $Q = 0.8$. The observation inclination angle of all the above images is $17°$. The black hole mass is normalized to $M = 1$

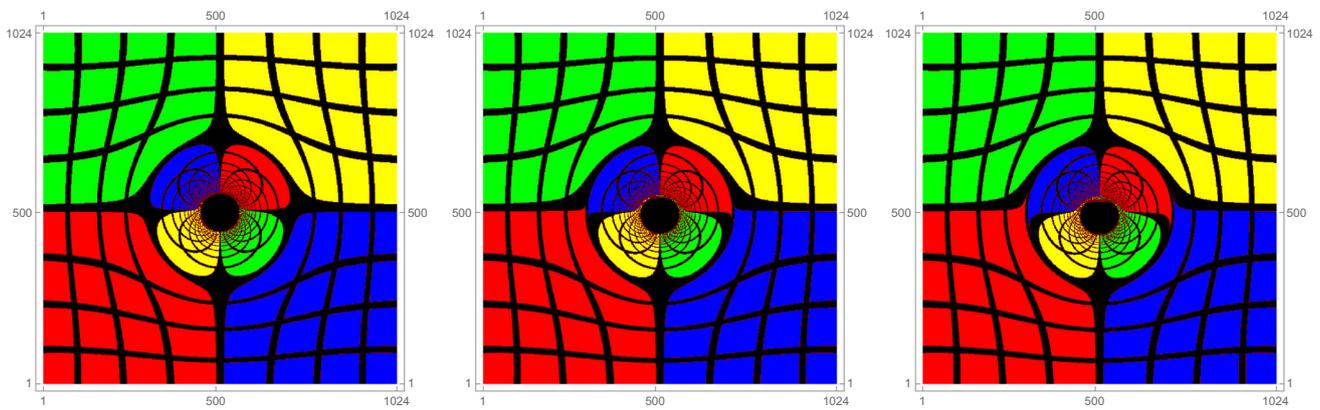

**Fig. 5** The shadow of the Kerr–Sen black hole is analyzed using the numerical ray-tracing method. From left to right, the parameters chosen are $a = 0.5$, $Q = 0.3$, $a = 0.8$, $Q = 0.3$, and $a = 0.95$, $Q = 0.3$. The observation inclination angle of all the above images is $17°$. The mass of the black hole is set to $M = 1$

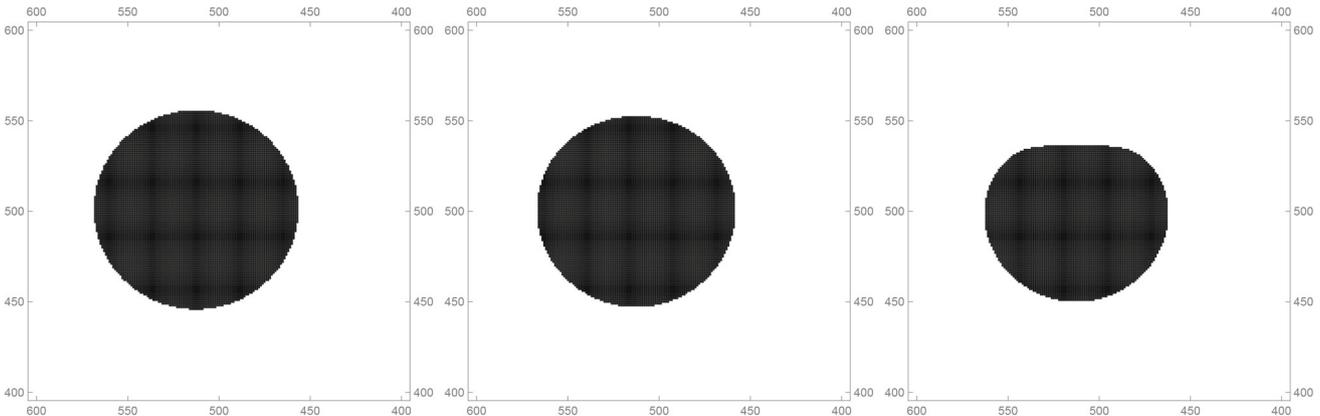

**Fig. 6** The inner shadow of the Kerr–Sen black hole. From left to right, the parameters are $a = 0.5$, $Q = 0.3$, $a = 0.5$, $Q = 0.5$, and $a = 0.5$, $Q = 0.8$. The observation inclination angle of all the above images is $17°$. The black hole mass is set to $M = 1$

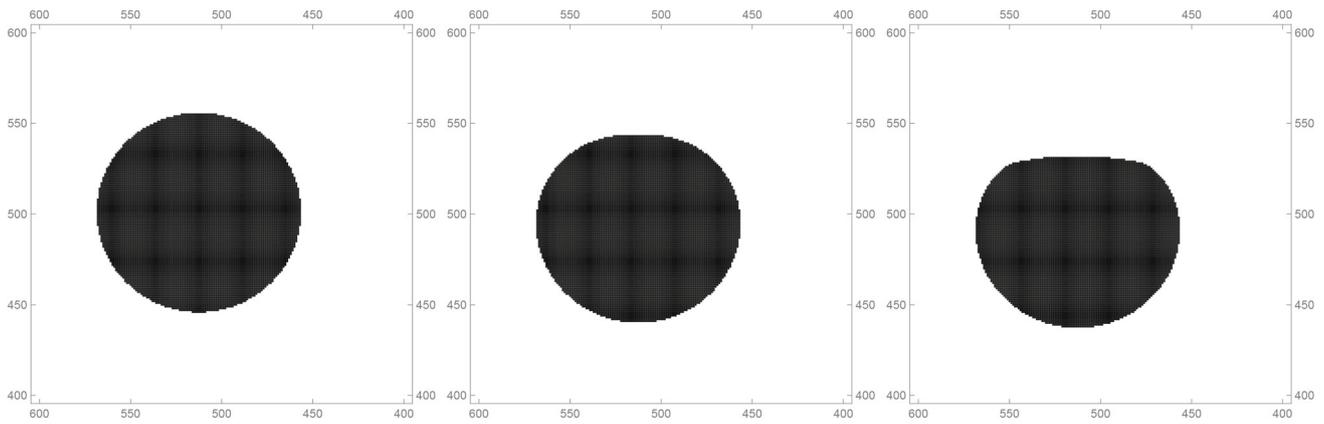

**Fig. 7** The inner shadow of the Kerr–Sen black hole. From left to right, the parameters are $a = 0.5$, $Q = 0.3$, $a = 0.8$, $Q = 0.3$, and $a = 0.95$, $Q = 0.3$. The observation inclination angle of all the above images is $17°$. The black hole mass is set to $M = 1$

**Fig. 8** A comparison of the inner shadow of the Kerr black hole. The parameters are $a = 0.5$, $Q = 0.3$, $a = 0.5$, $Q = 0.5$, and $a = 0.5$, $Q = 0.8$. The top row, from left to right, corresponds to $\theta = 17°$, $53°$, and the bottom row, from left to right, corresponds to $\theta = 75°$, $150°$. The mass of the black hole is set to $M = 1$

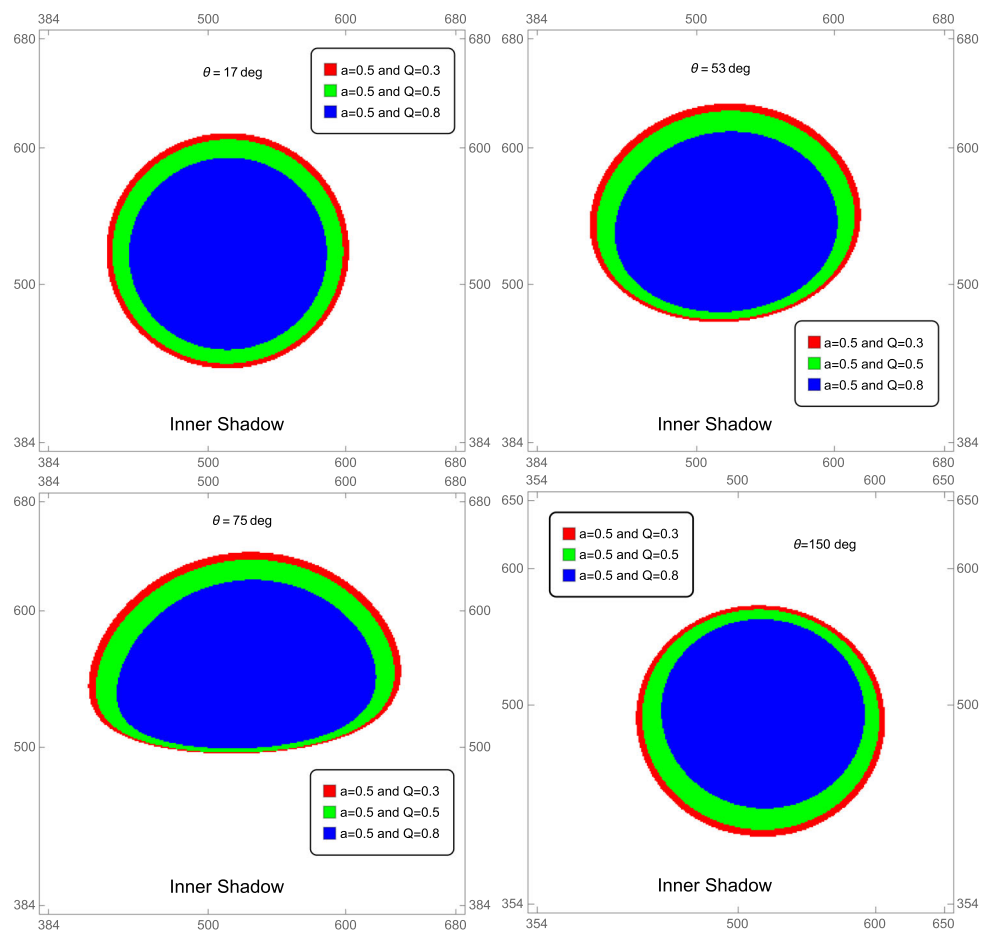

Here, we focus primarily on the radiation in the equatorial plane. At the position $\alpha, \beta$ on the image plane, we consider the number of intersections $N_{max}(\alpha, \beta)$ between the geodesics and the equatorial plane. When $N_{max} = 1$, the geodesic intersects the equatorial plane only once, projecting the direct image of the equatorial radiation onto the observer's sky. When $N_{max} = 2, 3, 4, \ldots$, lensing and higher-order images are produced.

Based on the focal points of the geodesics and their intersections with the equatorial plane, we can calculate the radius $r_m(\alpha, \beta)$ corresponding to the intersection between the geodesic passing through the position $(\alpha, \beta)$ on the observer's imaging plane and the equatorial plane. It is important to emphasize that the ray trajectory is determined by these intersection points with the equatorial plane rather than by the number of radial or angular turning points encountered along the path in spacetime. In Eq. (64), $J_{model}(r_m)$ denotes the emissivity at radius $r_m$ on the equatorial plane, $g$ is the redshift factor accounting for both gravitational and Doppler shifts, and $f_m$ is the fudge factor introduced to adjust the brightness of higher-order photon rings [47]. The accretion disk considered in our discussion is geometrically and optically thin; therefore, we set $f_m = 1.5$ here. The emissivity $J_{model}(r_m)$ can be expressed as [52]:

$$\log[J_{model}(r)] = A \log\left(\frac{r}{r_h}\right) + B \left(\log\left(\frac{r}{r_h}\right)\right)^2. \quad (65)$$

**Fig. 9** A comparison of the inner shadow of the Kerr black hole. The parameters are $a = 0.5$, $Q = 0.3$, $a = 0.8$, $Q = 0.3$, and $a = 0.95$, $Q = 0.3$. The top row, from left to right, corresponds to $\theta = 17°$, $53°$, and the bottom row, from left to right, corresponds to $\theta = 75°$, $150°$. The mass of the black hole is set to $M = 1$

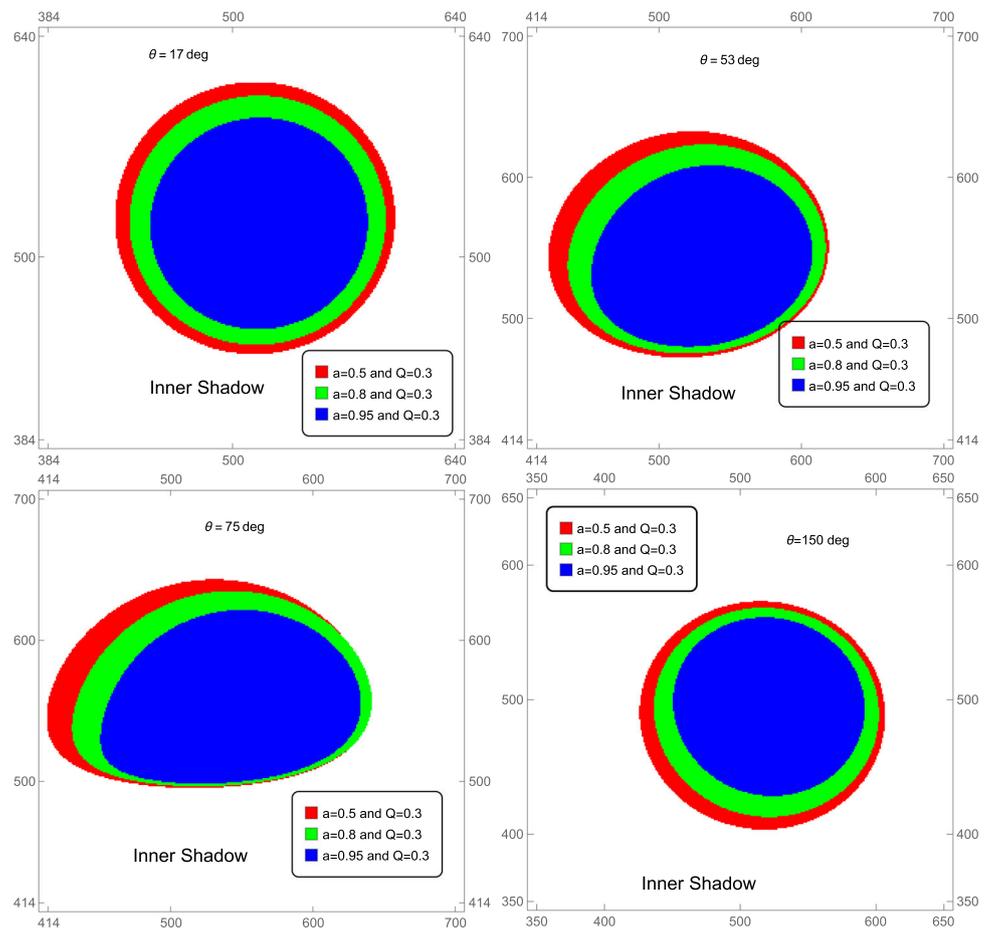

At an observational frequency of 230 GHz, the corresponding wavelength for M87* and Sgr A* is approximately 1.3 mm. We adopt the parameters $A = -2$ and $B = -\frac{1}{2}$ for this frequency. When the observation frequency shifts to 86 GHz, these parameters are updated to $A = 0$ and $B = -\frac{3}{4}$ [53]. This framework enables the calculation of the luminosity of the Kerr–Sen black hole within the thin disk model.

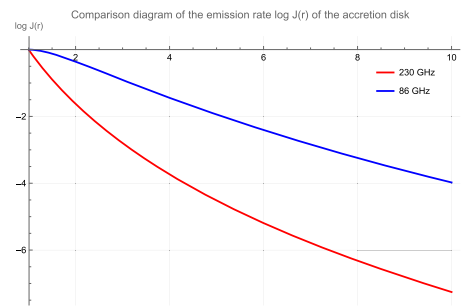

**Fig. 10** A comparison diagram of the emission rate log J(r) of the accretion disk under the conditions of 230 GHz and 86 GHz. The blue curve represents the emissivity profile of the accretion disk at 86 GHz, with parameters $A = 0$, $B = -\frac{3}{4}$. The red curve corresponds to the 230 GHz case, where $A = -2$, $B = -\frac{1}{2}$. The horizon radius is fixed at $r_H = 1$

For the redshift factor $g = \frac{\nu_o}{\nu_m}$, since the accretion disk model employed here extends beyond previous approximations by allowing photons emitted from the inner edge of the disk to approach the black hole's event horizon, it permits a more accurate investigation of redshift effects [28].

Outside the ISCO, the accretion flow continues along stable circular orbits characterized by an angular velocity given by $\Omega_m(r) = \frac{\mu^\phi}{\mu^r}\big|_{r=r_m}$. The parameter $\xi$ is defined by Eq. (6), and $\epsilon$ denotes the ratio of the energy measured on the observer's screen to the conserved energy along the geodesic [14]:

$$\epsilon = \frac{E_0}{E} = \frac{P_t}{k_t} = \delta(1 + \xi\chi). \tag{66}$$

Here, $\chi$ and $\delta$ are defined by Eq. (58). In asymptotically flat spacetime, $\epsilon = 1$, but in the non-asymptotically flat Kerr–Sen black hole spacetime, $\epsilon$ is less than 1. Therefore, when selecting the position of $r_m$, if it exceeds the ISCO, the redshift factor is expressed as

$$g_m = \frac{\epsilon}{\zeta(1 - \xi\Omega_m)}, \quad r_m > r_{\mathrm{ISCO}}. \tag{67}$$

where

$$\zeta = \sqrt{-\frac{1}{g_{tt} + 2g_{t\phi}\Omega_m + g_{\phi\phi}\Omega_m^2}}\Bigg|_{r=r_m}. \tag{68}$$

As previously mentioned, the ISCO defines the boundary of the accretion disk region. Within the accretion disk region, the accretion flow moves along plunging orbits with a radial velocity $u_c^r$. In this case, the redshift factor can be expressed as:

$$g_m = \frac{\epsilon}{-\frac{u_c^r k_r}{k_t} + g^{tt}E_{\mathrm{ISCO}} - \xi g^{t\phi}E_{\mathrm{ISCO}} - g^{t\phi}L_{\mathrm{ISCO}} + \xi g^{\phi\phi}L_{\mathrm{ISCO}}}. \tag{69}$$

To more intuitively depict the redshift, we fix the observation radius at $r_o = 100$, and set the observation angle on the image plane to $\phi = \frac{\pi}{10}$. Furthermore, we consider both prograde and retrograde motions of the accretion disk, distinguishing photons emitted in the forward and backward directions relative to the black hole's rotation. This distinction is crucial, as the direction of the accretion disk's rotation significantly influences both the observed redshift and the photon trajectories around the Kerr–Sen black hole (Fig. 10).

Figures 11, 12, 13, 14, 15, 16, 17 and 18 display the direct and lensing redshift distribution maps, depicting the redshift profiles of the accretion disk in both prograde and retrograde configurations. It is clearly observed that as the observational inclination angle $\theta_0$ increases from 0° to 90°, the redshifted region systematically shrinks, while the blueshifted region correspondingly expands. Conversely, as the inclination angle $\theta_0$ increases from 90° to 180°, the redshifted region gradually enlarges, and the blueshifted region contracts.

The diagrams also depict the redshift distributions of the accretion disk for varying magnitudes of charge and spin parameters. It is well known that as the spin parameter increases, the ISCO radius moves closer to the black hole, causing particles to release energy nearer to the event horizon, which leads to stronger gravitational redshifts. Similarly, increasing the charge magnitude induces effects analogous to those produced by higher spin parameters, further enhancing the gravitational redshift. However, as shown in the diagrams, variations in the location and magnitude of the maximum blueshift due to changes in charge and spin parameters remain minimal, and are far less pronounced than those induced by changes in the observation inclination angle. Consequently, we conclude that the inclination angle of observation is the dominant factor influencing the variations in redshift and blueshift distributions, while the effects of the black hole's spin and charge parameters are comparatively minor.

### 3.3 Image of the Keer–Sen black hole within a thin disk

Next, we will focus on the images of the Kerr–Sen black hole. Using Eq. (63) and the fisheye camera ray-tracing technique, we can generate images of the Kerr–Sen black hole illuminated by a thin accretion disk. For a frequency of 230 GHz, we select parameters $A = -2$ and $B = -\frac{1}{2}$, which gives the following formula:

$$\log[J_{\mathrm{model}}(r)] = -2\log\left(\frac{r}{r_h}\right) - \frac{1}{2}\left(\log\left(\frac{r}{r_h}\right)\right)^2. \tag{70}$$

To achieve a more comprehensive understanding of the observational appearance of the Kerr–Sen black hole, we have also computed the results at 86 GHz and conducted a comparative analysis with observations at 230 GHz. Based on Eqs. (69) and (70), we first present the intensity distribution along the $x$- and $y$-axes. Figures 19 and 20 display the intensity profiles of the accretion disk for both prograde and retrograde motion along the $x$- and $y$-axes, respectively.

Along the $x$-axis, the intensity peak decreases irrespective of whether the accretion disk exhibits prograde or retrograde motion. The spin parameter induces a comparable effect, which is especially pronounced for the retrograde accretion disk. Variations in the black hole's spin parameter result in a contraction of the central intensity region toward the black hole. Concerning the observer's inclination angle, for a prograde accretion disk, an increase in the angle corresponds to a gradual decrease in the intensity peak and an inward contraction of the central intensity. In contrast, for a retrograde accretion disk, the intensity peak initially increases before subsequently decreasing as the inclination angle grows, reaching a maximum at $\theta = 53°$. Correspondingly, the central intensity first contracts inward and then expands outward.

Along the $y$-axis, the influences of the charge and spin parameters are comparable to those observed along the $x$-axis. However, irrespective of whether the accretion disk exhibits prograde or retrograde motion, the intensity peaks transition from being roughly symmetric to exhibiting pronounced asymmetry. In the prograde case, the intensity on the left side is markedly higher than on the right, whereas in the retrograde case, this asymmetry is reversed. Furthermore,

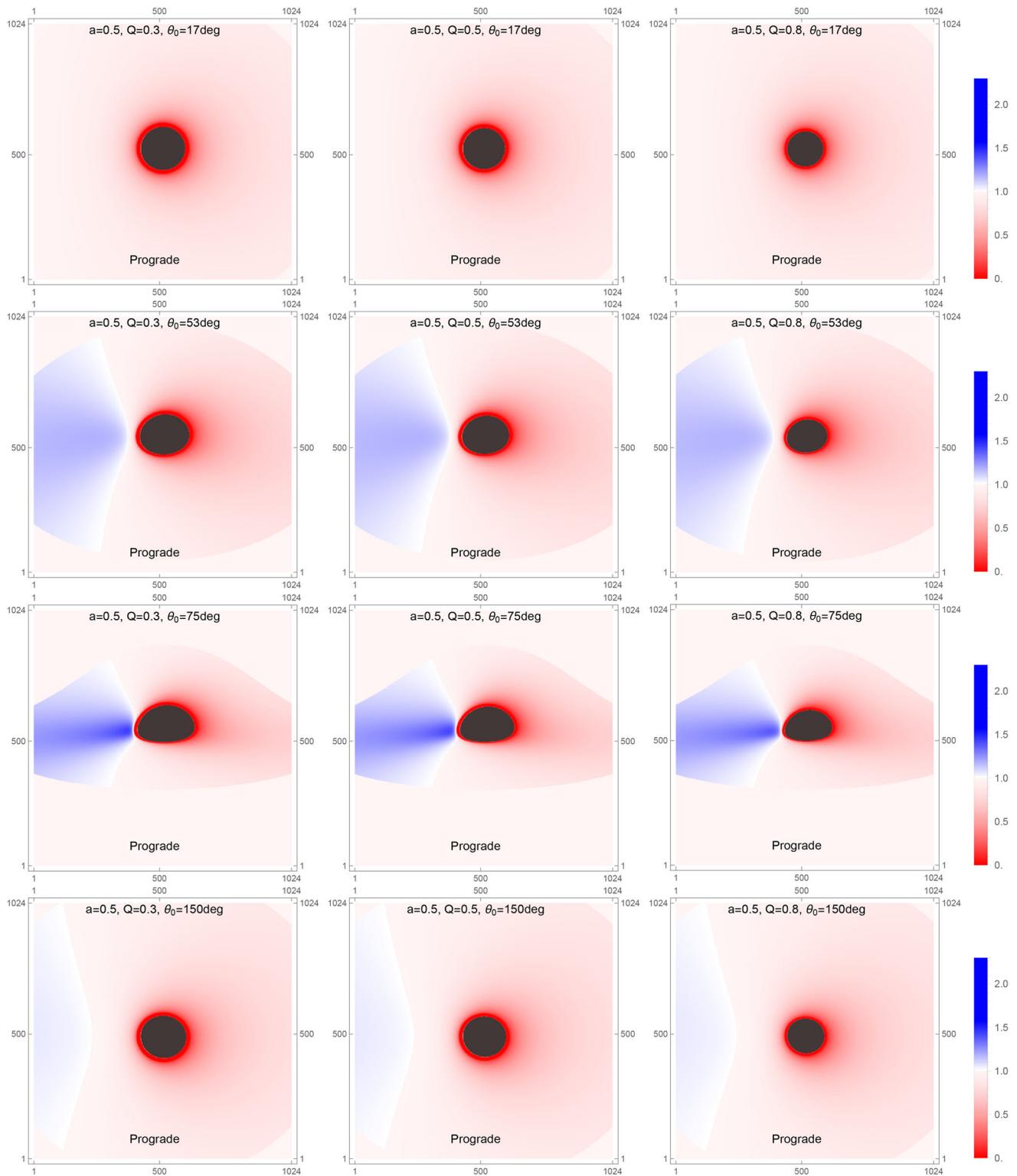

**Fig. 11** First-order redshift image. The above images show the prograde redshift distribution for different parameters of the accretion disk and observation angles. We selected three sets of data: $a = 0.5$, $Q = 0.3$, $a = 0.5$, $Q = 0.5$, and $a = 0.5$, $Q = 0.8$, with different electric charges. These sets were compared at four observation angles: $17°$, $53°$, $75°$, and $150°$. The black hole mass is set to $M = 1$

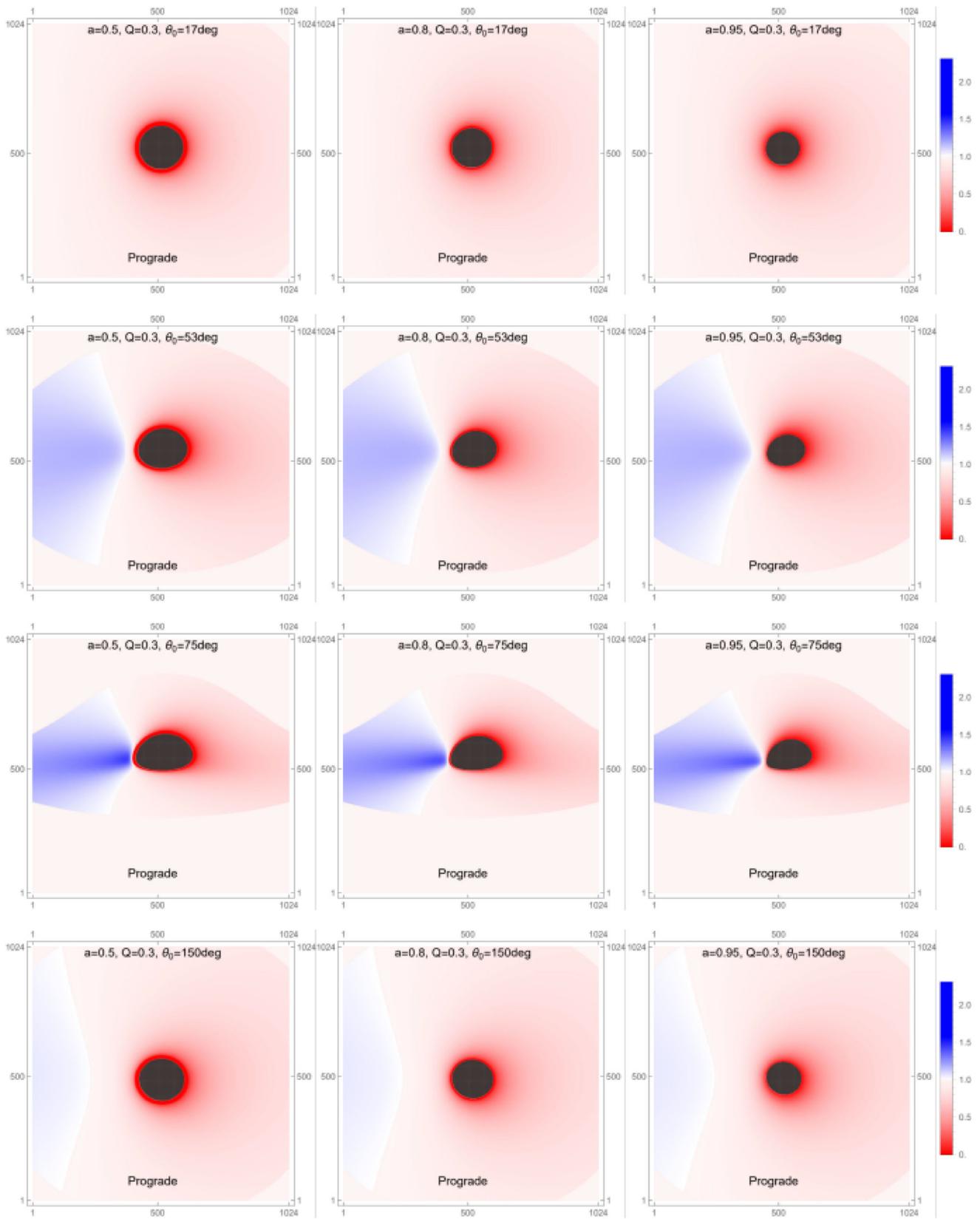

**Fig. 12** First-order redshift image. The above images show the prograde redshift distribution for different parameters of the accretion disk and observation angles. We selected three sets of data: $a = 0.5$, $Q = 0.3$, $a = 0.8$, $Q = 0.3$, and $a = 0.95$, $Q = 0.3$, with different spin parameters. These sets were compared at four observation angles: $17°$, $53°$, $75°$, and $150°$. The black hole mass is set to $M = 1$

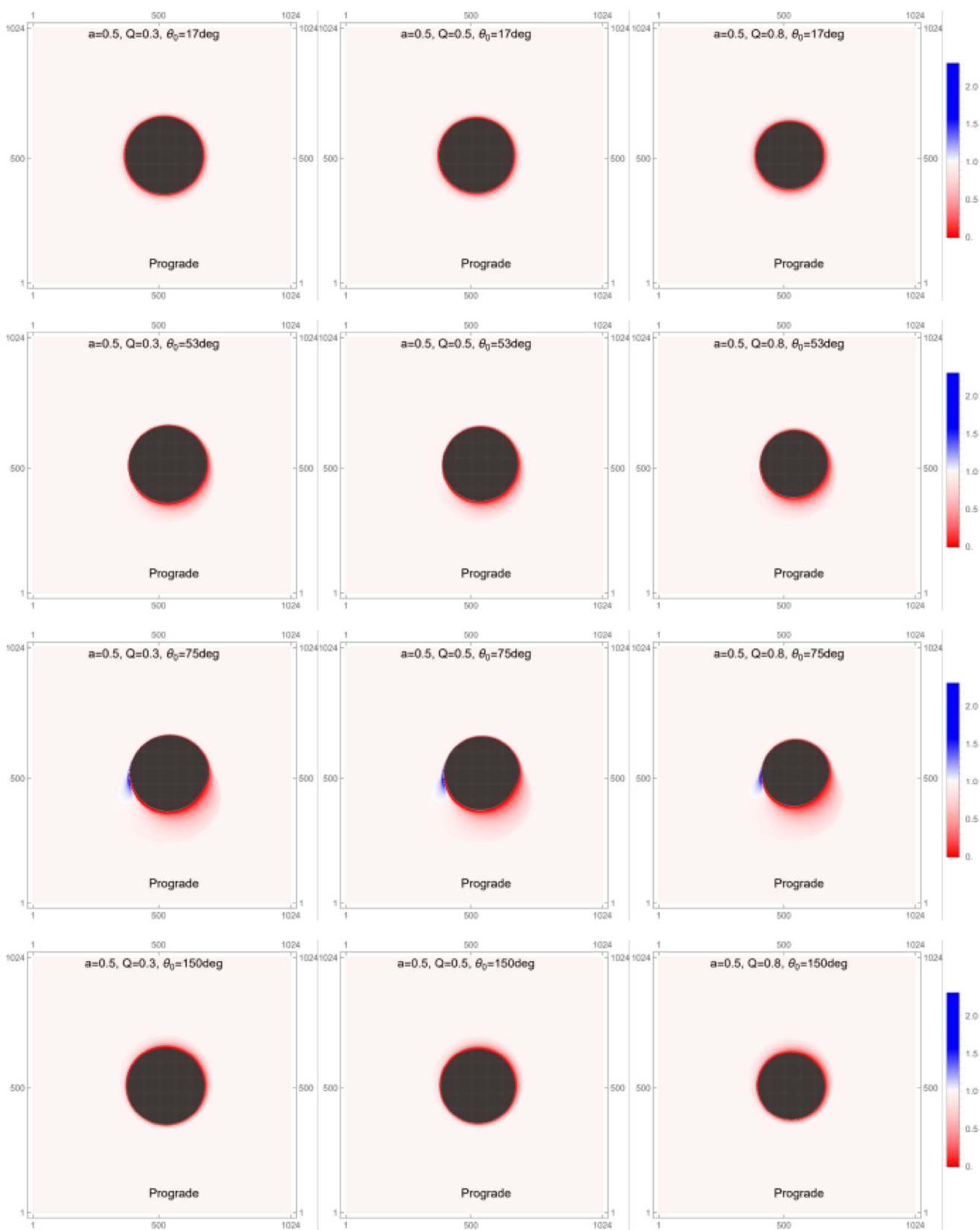

**Fig. 13** Second-order redshift image. The above images show the prograde redshift distribution for different parameters of the accretion disk and observation angles. We selected three sets of data: $a = 0.5$, $Q = 0.3$, $a = 0.5$, $Q = 0.5$, and $a = 0.5$, $Q = 0.8$, with different electric charges. These sets were compared at four observation angles: 17°, 53°, 75°, and 150°. The black hole mass is set to $M = 1$

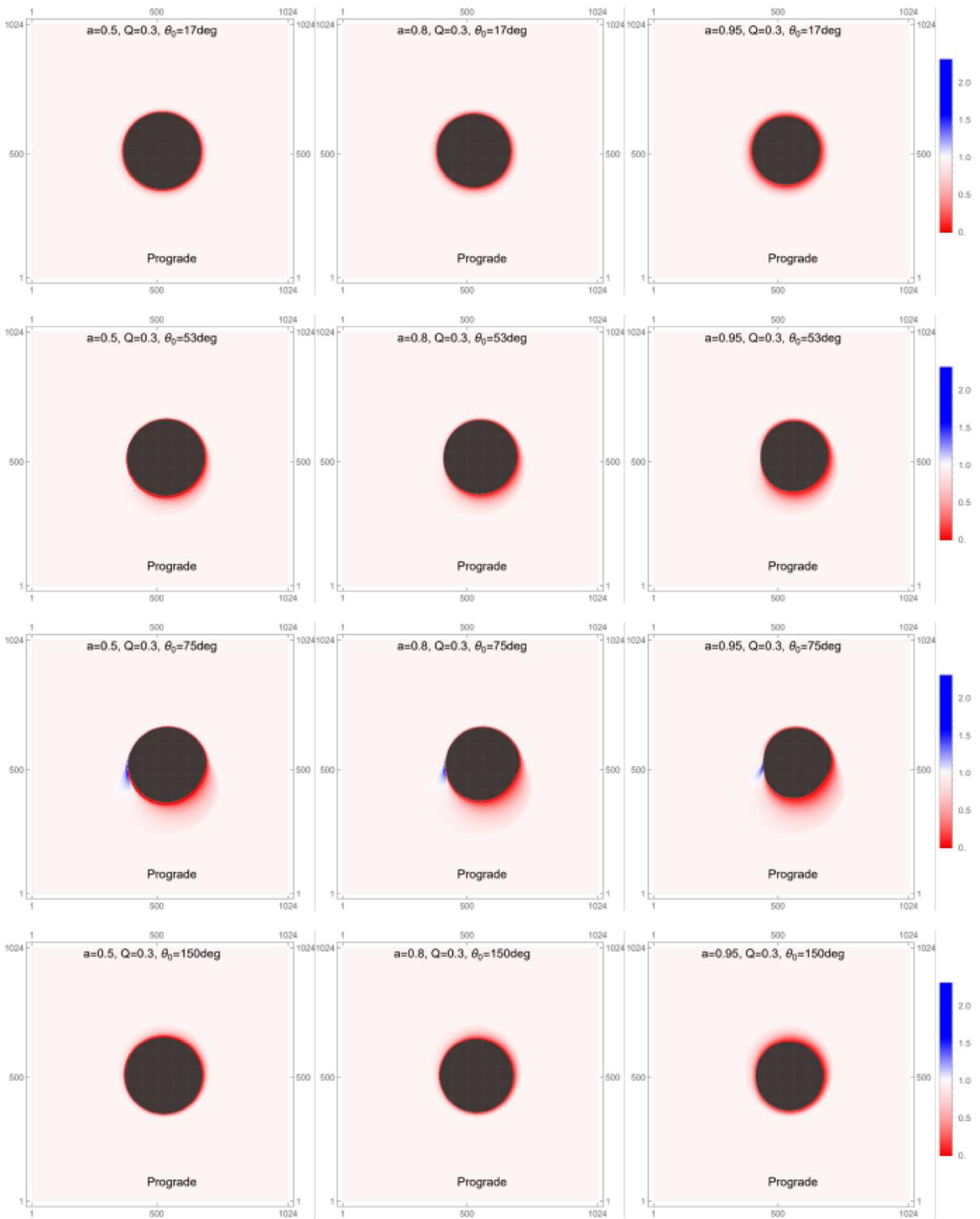

**Fig. 14** Second-order redshift image. The above images show the prograde redshift distribution for different parameters of the accretion disk and observation angles. We selected three sets of data: $a = 0.5$, $Q = 0.3$, $a = 0.8$, $Q = 0.3$, and $a = 0.95$, $Q = 0.3$, with different spin parameters. These sets were compared at four observation angles: 17°, 53°, 75°, and 150°. The black hole mass is set to $M = 1$

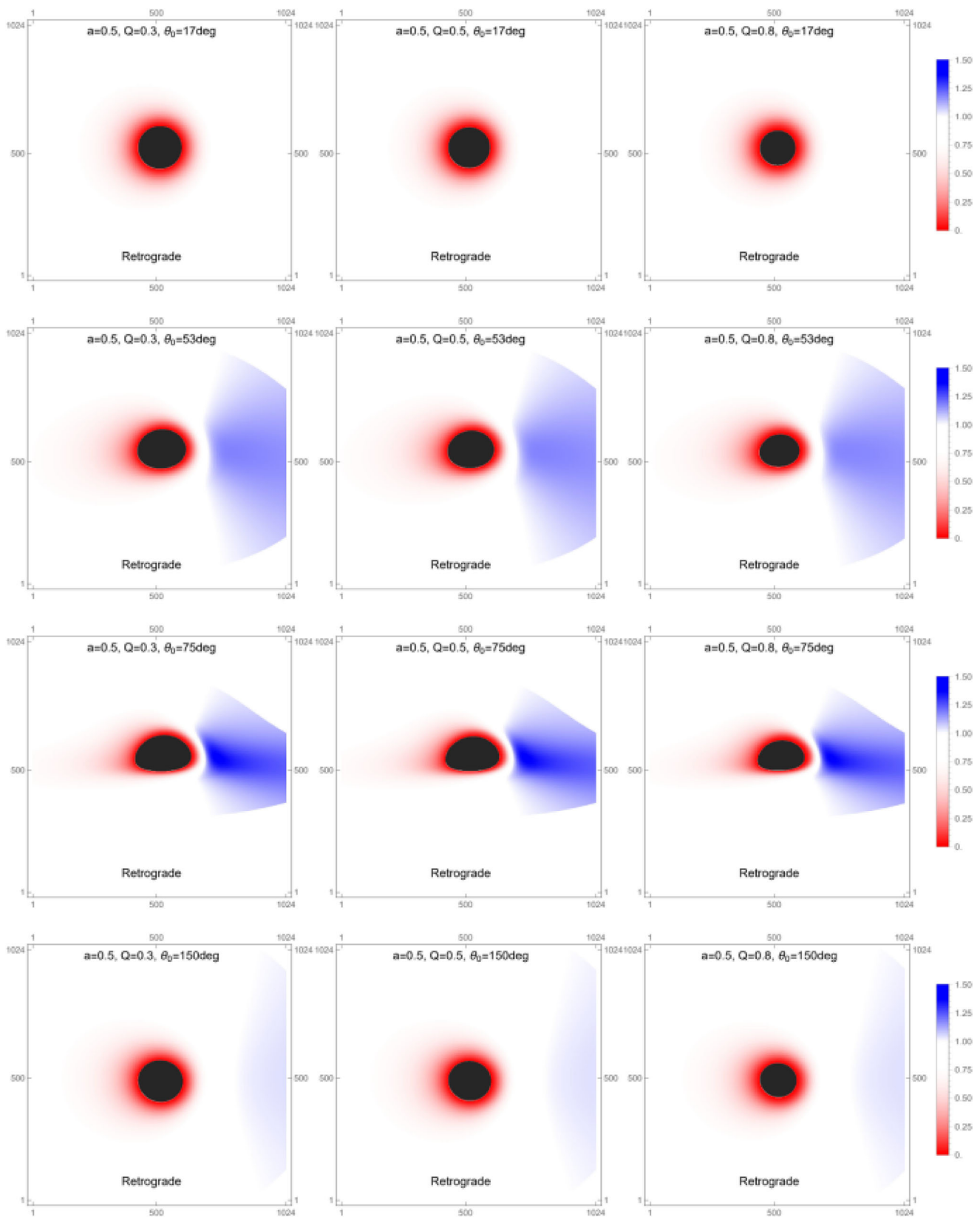

**Fig. 15** First-order redshift image. The above images show the retrograde redshift distribution for different parameters of the accretion disk and observation angles. We selected three sets of data: $a = 0.5$, $Q = 0.3$, $a = 0.5$, $Q = 0.5$, and $a = 0.5$, $Q = 0.8$, with different electric charges. These sets were compared at four observation angles: $17°$, $53°$, $75°$, and $150°$. The black hole mass is set to $M = 1$

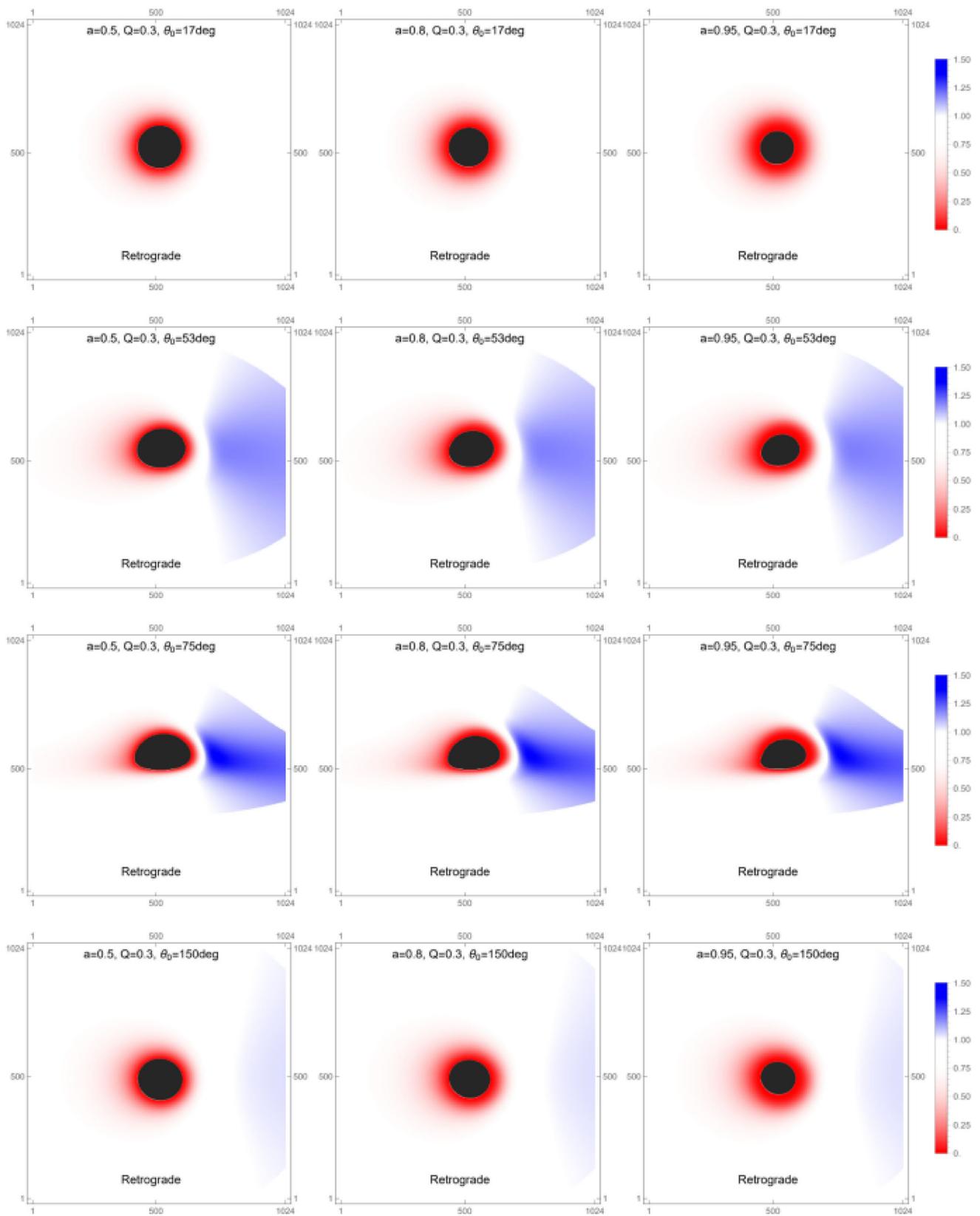

**Fig. 16** First-order redshift image. The above images show the retrograde redshift distribution for different parameters of the accretion disk and observation angles. We selected three sets of data: $a = 0.5$, $Q = 0.3$, $a = 0.8$, $Q = 0.3$, and $a = 0.95$, $Q = 0.3$, with different spin parameters. These sets were compared at four observation angles: $17°$, $53°$, $75°$, and $150°$. The black hole mass is set to $M = 1$

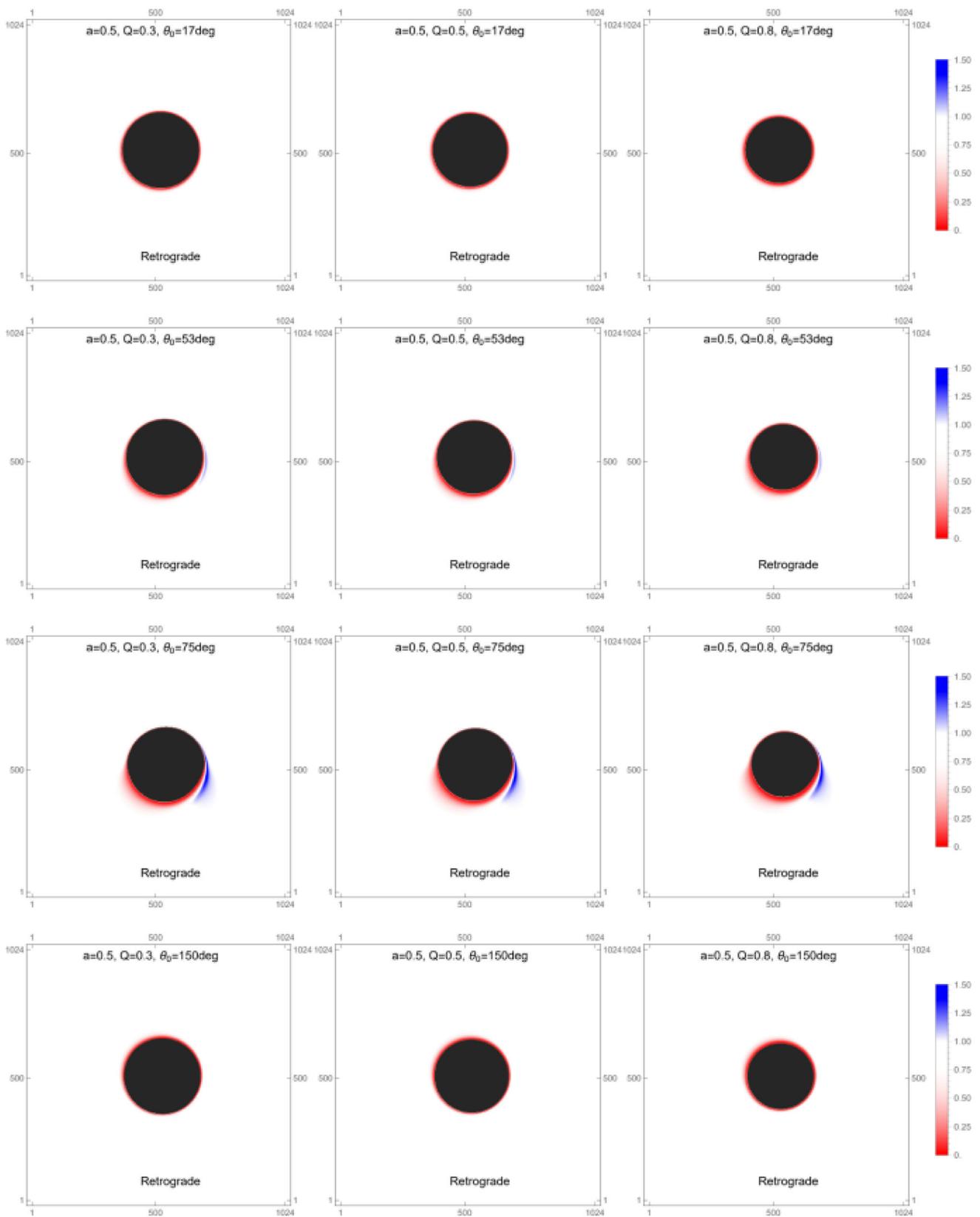

**Fig. 17** Second-order redshift image. The above images show the retrograde redshift distribution for different parameters of the accretion disk and observation angles. We selected three sets of data: $a = 0.5$, $Q = 0.3$, $a = 0.5$, $Q = 0.5$, and $a = 0.5$, $Q = 0.8$, with different electric charges. These sets were compared at four observation angles: $17°$, $53°$, $75°$, and $150°$. The black hole mass is set to $M = 1$

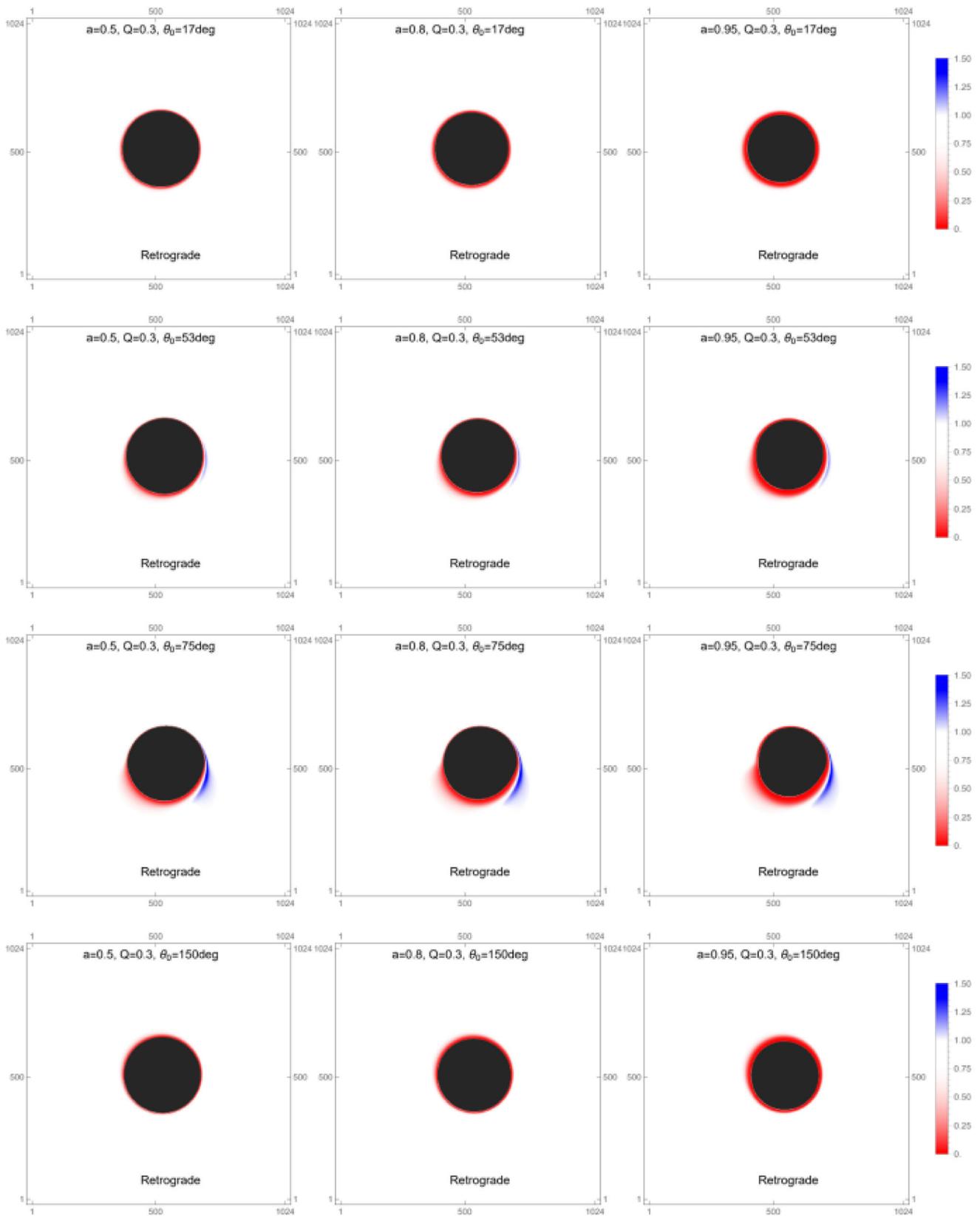

**Fig. 18** Second-order redshift image. The above images show the retrograde redshift distribution for different parameters of the accretion disk and observation angles. We selected three sets of data: $a = 0.5$, $Q = 0.3$, $a = 0.8$, $Q = 0.3$, and $a = 0.95$, $Q = 0.3$, with different spin parameters. These sets were compared at four observation angles: 17°, 53°, 75°, and 150°. The black hole mass is set to $M = 1$

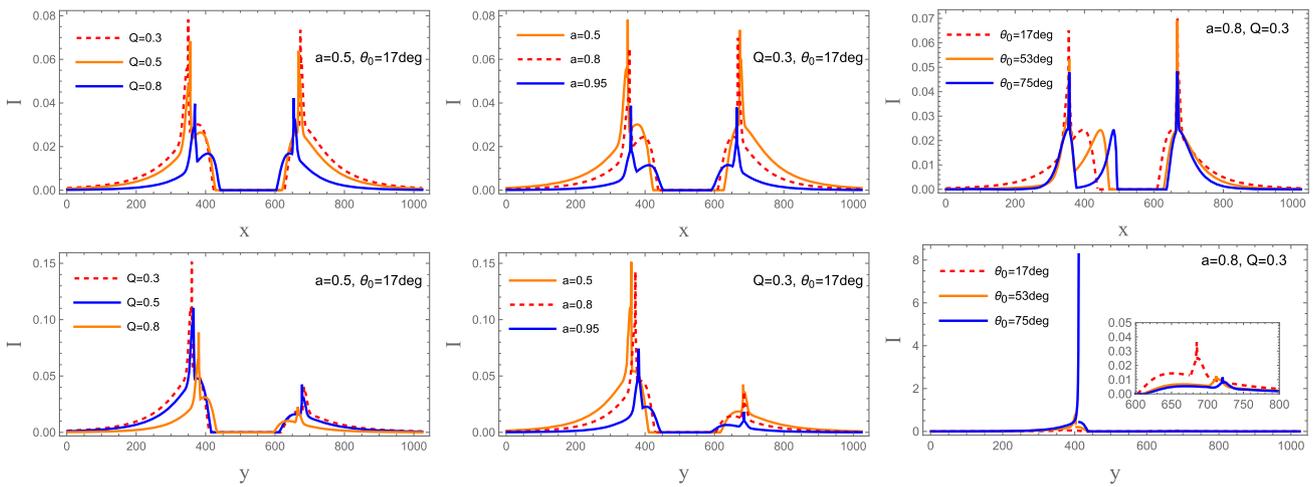

**Fig. 19** Image of the intensity comparison. The intensity distribution of the Kerr–Sen black hole at 230 GHz when the accretion disk is moving in the prograde direction. First row: intensity distribution along the $x$-axis. Second row: intensity distribution along the $y$-axis

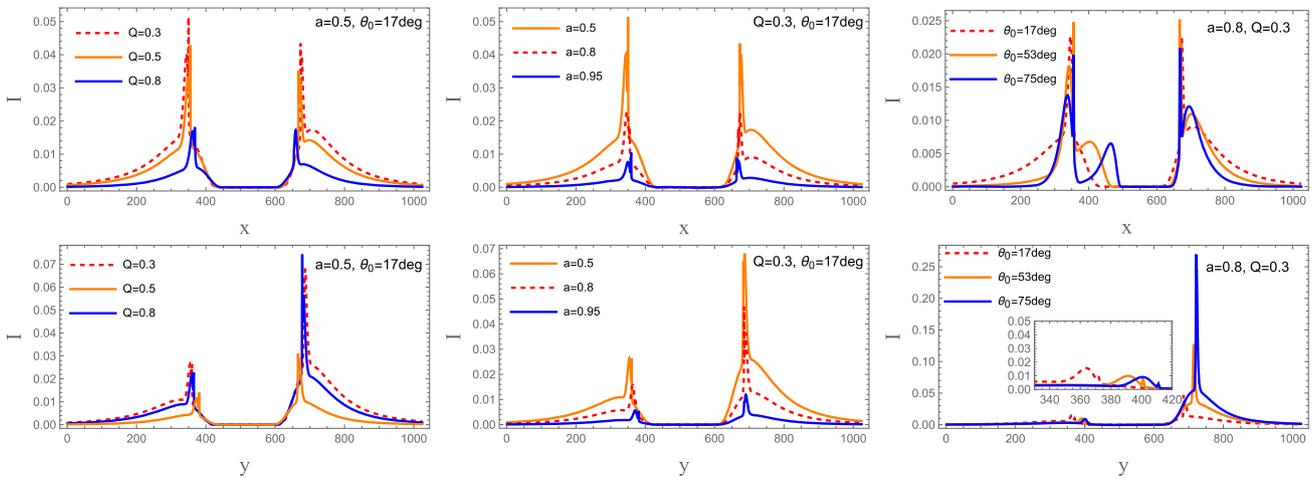

**Fig. 20** Image of the intensity comparison. The intensity distribution of the Kerr–Sen black hole at 230 GHz when the accretion disk is moving in the retrograde direction. First row: intensity distribution along the $x$-axis. Second row: intensity distribution along the $y$-axis

with increasing observer inclination angle, the intensity peak increases sharply, accompanied by a shift of the peak position toward the right.

Figures 21, 22, 23 and 24 display images of the Kerr–Sen black hole surrounded by a thin accretion disk, considering both prograde and retrograde disk motions. Regardless of the disk's rotational direction, for $\theta < \frac{\pi}{2}$, the direct and lensed images become progressively more distinguishable as the inclination angle increases. However, at an observational angle of $\theta = 150°$, the reduced brightness renders the distinction between the direct and lensed images increasingly challenging.

In addition, the black hole shadow of the Kerr–Sen black hole is influenced by both the electric charge and the spin parameter, with the impact of the spin parameter on the shadow being substantially more pronounced than that of the charge. However, irrespective of variations in the black hole's spin and charge, the inner shadow and the critical curve remain well-defined in both low- and high-inclination observations, indicating that these features are intrinsic to the black hole's spacetime geometry. In scenarios where the accretion disk is geometrically thin, the decrease in central brightness also emerges as a characteristic signature. Furthermore, the Doppler effect induced by both prograde and retrograde motions is clearly manifested on opposite sides of the image plane. For retrograde accretion disks, the Doppler effect is enhanced on the right side of the imaging screen, although the overall brightness of the Kerr–Sen black hole is diminished due to the rotational frame-dragging effect. Beyond the images obtained at a frequency of 230 GHz, we also present images of the Kerr–Sen black hole surrounded by a thin accretion disk at 86 GHz for comparative analysis. Here, Eq. (69) is transformed [53]:

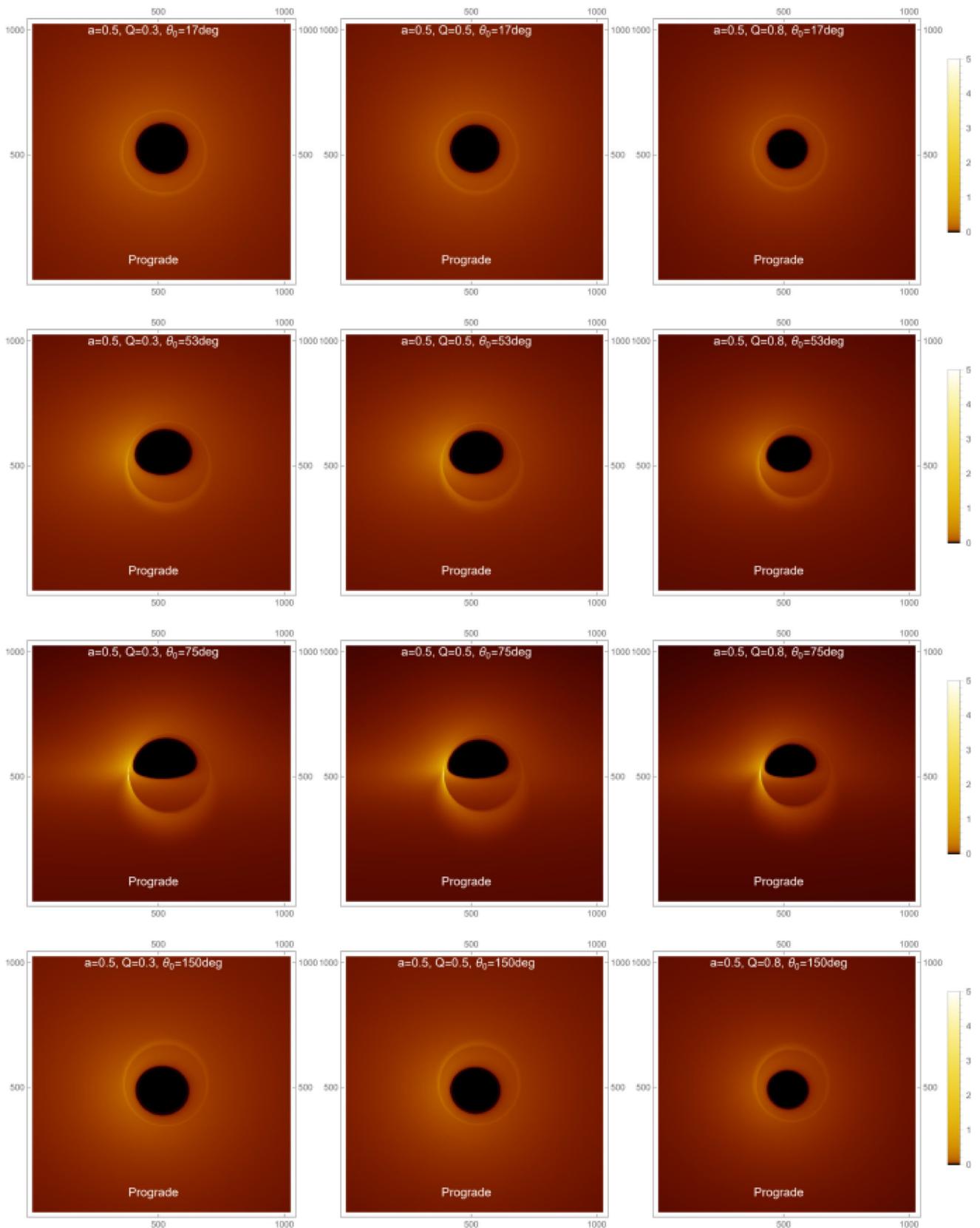

**Fig. 21** The image of the Kerr–Sen black hole surrounded by a prograde accretion disk at 230 GHz. From left to right: $a = 0.5$, $Q = 0.3$, $a = 0.5$, $Q = 0.5$, and $a = 0.5$, $Q = 0.8$. From top to bottom: 17°, 53°, 75°, and 150°, with the black hole mass set to $M = 1$

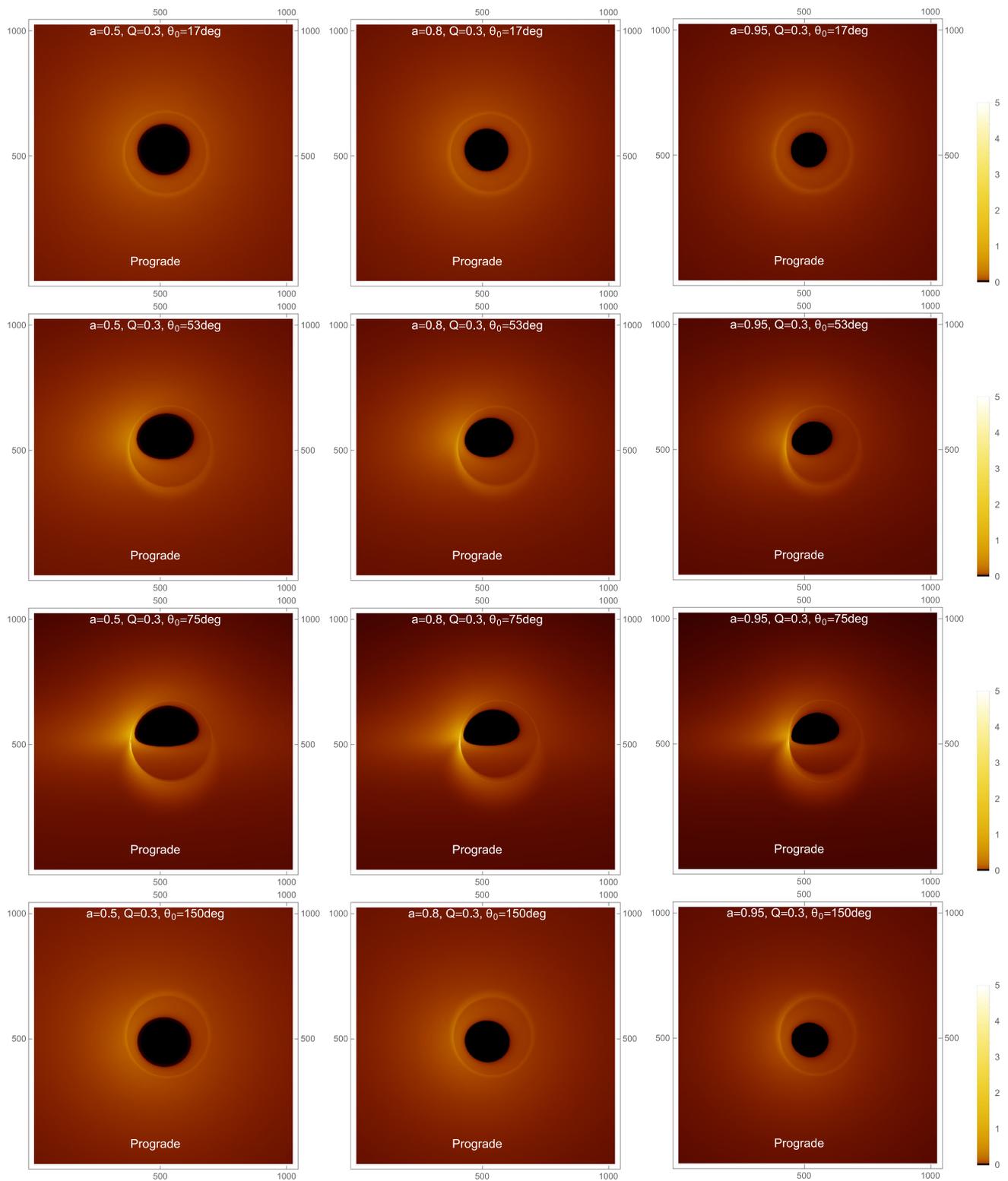

**Fig. 22** The image of the Kerr–Sen black hole surrounded by a prograde accretion disk at 230 GHz. From left to right: $a = 0.5$, $Q = 0.3$, $a = 0.8$, $Q = 0.3$, and $a = 0.95$, $Q = 0.3$. From top to bottom: $17°$, $53°$, $75°$, and $150°$, with the black hole mass set to $M = 1$

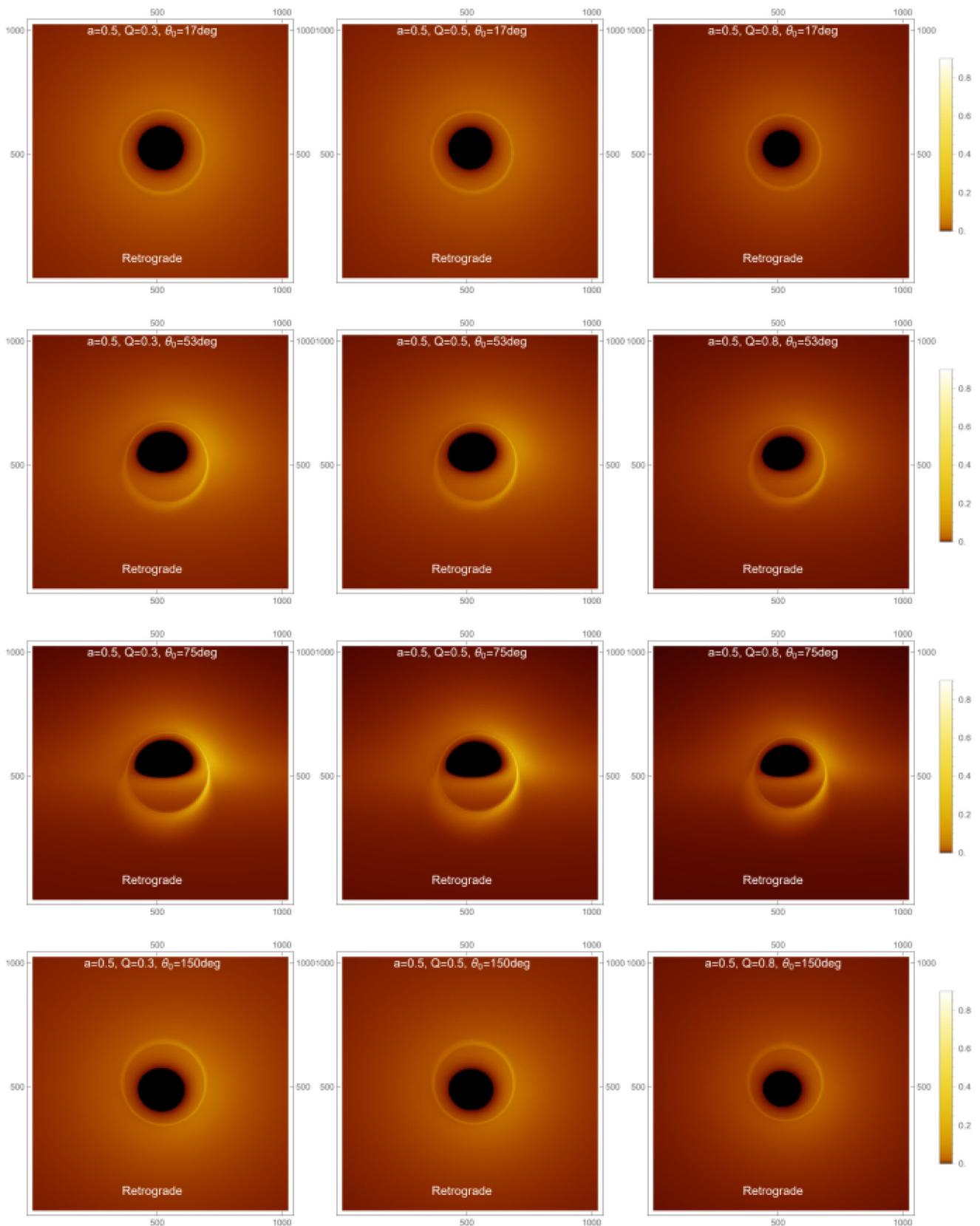

**Fig. 23** The image of the Kerr–Sen black hole surrounded by a retrograde accretion disk at 230 GHz. From left to right: $a = 0.5$, $Q = 0.3$, $a = 0.5$, $Q = 0.5$, and $a = 0.5$, $Q = 0.8$. From top to bottom: 17°, 53°, 75°, and 150°, with the black hole mass set to $M = 1$

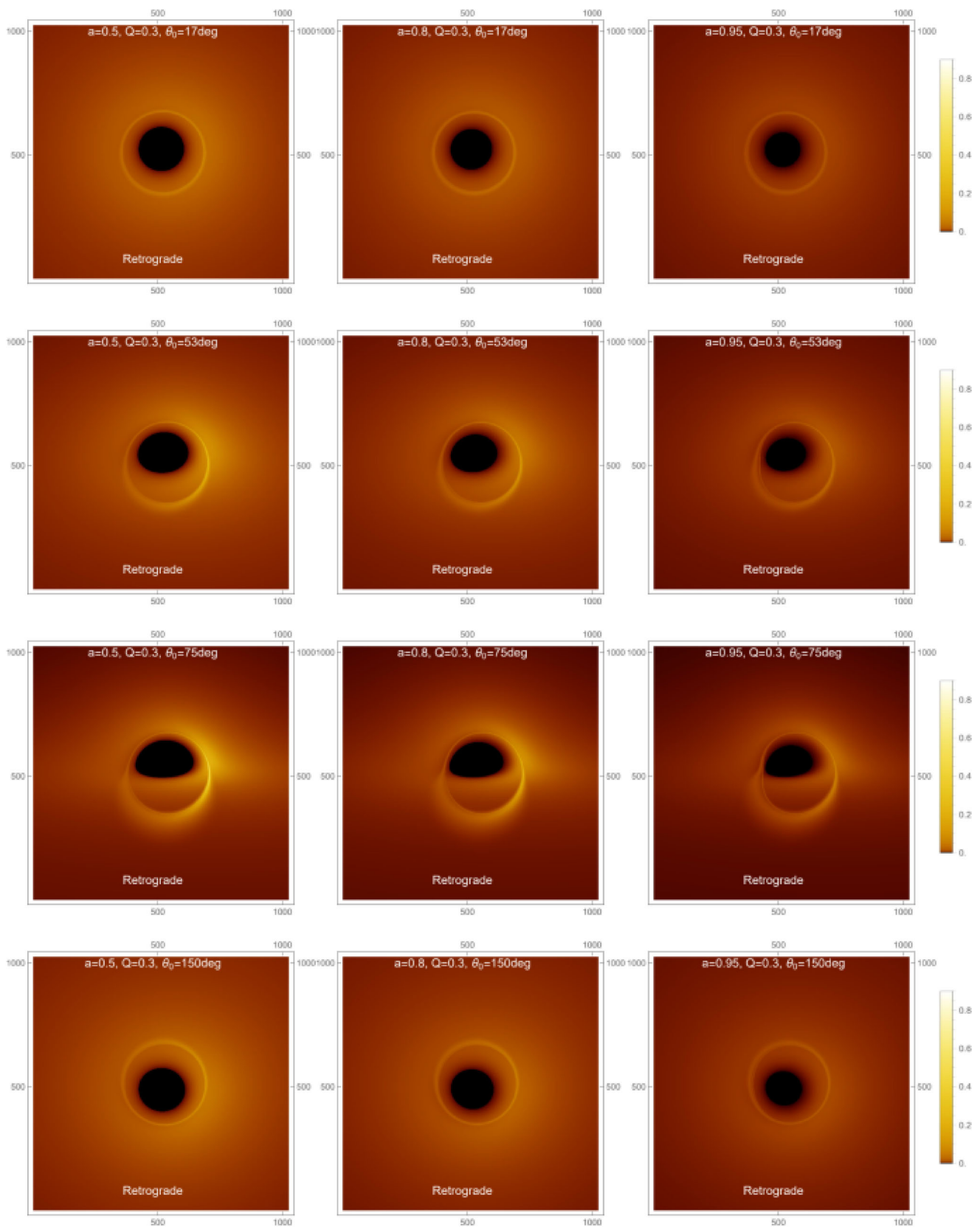

**Fig. 24** The image of the Kerr–Sen black hole surrounded by a retrograde accretion disk at 230 GHz. From left to right: $a = 0.5$, $Q = 0.3$, $a = 0.8$, $Q = 0.3$, and $a = 0.95$, $Q = 0.3$. From top to bottom: $17°$, $53°$, $75°$, and $150°$, with the black hole mass set to $M = 1$

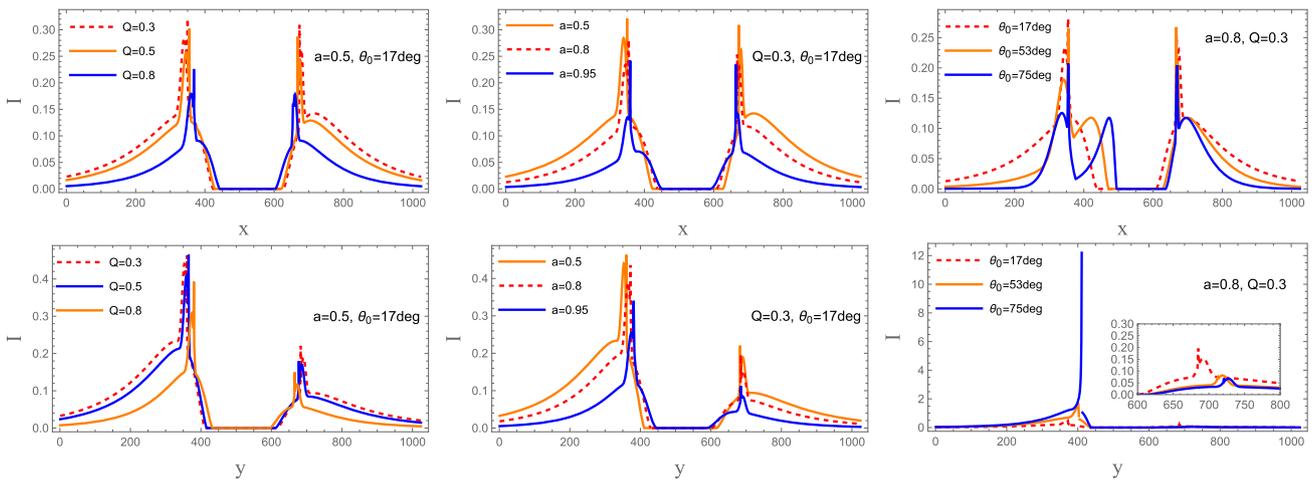

**Fig. 25** Intensity distribution of the Kerr–Sen black hole surrounded by a prograde accretion disk at 86 GHz. The first row shows the intensity distribution along the x-axis, while the second row shows the intensity distribution along the y-axis

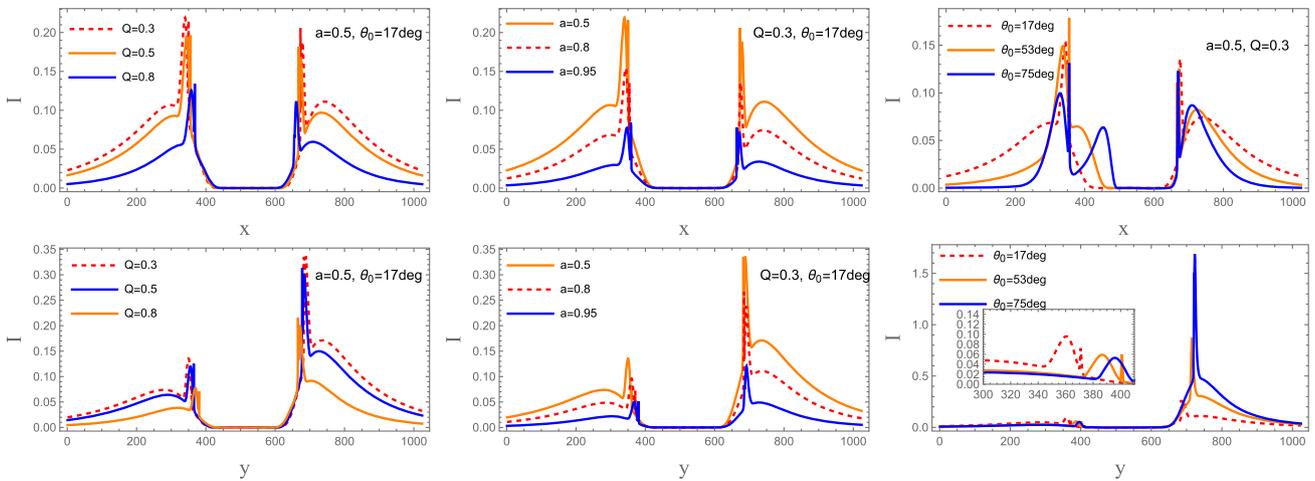

**Fig. 26** Intensity distribution of the Kerr–Sen black hole surrounded by a retrograde accretion disk at 86 GHz. The first row shows the intensity distribution along the x-axis, while the second row shows the intensity distribution along the y-axis

$$\log[J_{model}(r)] = -\frac{3}{4}\left(\log\left[\frac{r}{r_h}\right]\right)^2, \tag{71}$$

Figures 25 and 26 present the intensity distributions along the *x*-axis and *y*-axis for the Kerr–Sen black hole surrounded by a thin accretion disk at 86 GHz. Compared to the 230 GHz case, the overall intensity at 86 GHz is substantially higher, irrespective of whether the accretion disk is in prograde or retrograde motion. Although there is some variation in the position of the maximum peak, the influence of the charge, spin parameter, and observation angle on the intensity is qualitatively consistent with that observed at 230 GHz. Finally, we produce an image of the Kerr–Sen black hole with a thin accretion disk at 86 GHz and compare it with the corresponding scene at 230 GHz. The radiation brightness within the field of view is notably affected by gravitational redshift, while the central brightness depression, or inner shadow, remains prominent. Additionally, the critical curve remains unchanged, indicating that the inner shadow is an intrinsic feature of the black hole's spacetime (Figs. 27, 28, 29, 30).

## 4 Conclusion and discussion

In this study, we have conducted a thorough investigation into the observable properties of a Kerr–Sen black hole surrounded by a thin accretion disk. By employing ray-tracing techniques and analyzing photon trajectories, we systematically examined the formation of the black hole's image and its dependence on the black hole's charge and spin parameters. Through the application of elliptic integrals, we derived a fourth-order equation to determine the critical curve and the inner shadow. Our results demonstrate that both the spin

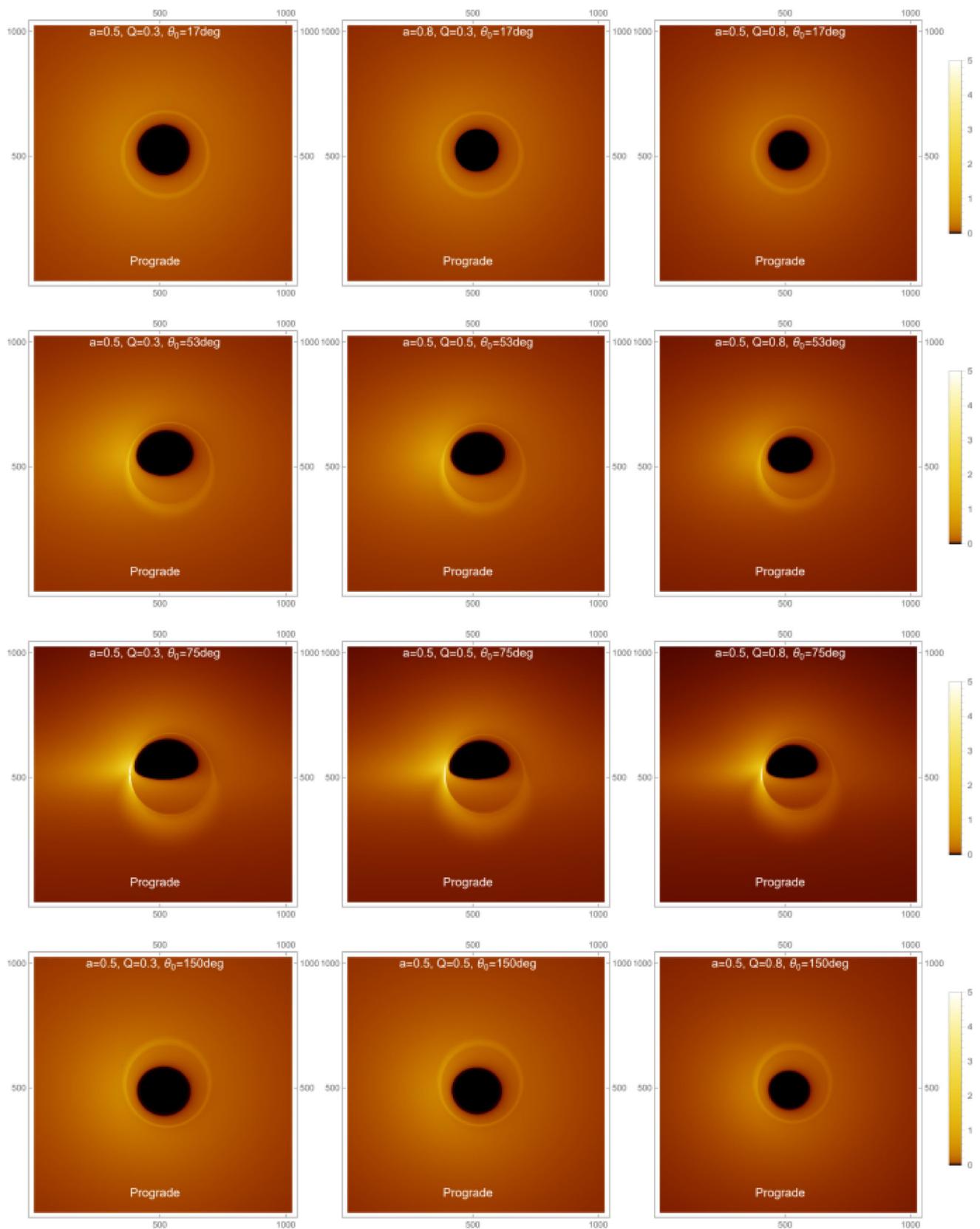

**Fig. 27** Images of the Kerr–Sen black hole surrounded by a prograde accretion disk at 86 GHz. From left to right, the parameters are: $a = 0.5$, $Q = 0.3$, $a = 0.5$, $Q = 0.5$, and $a = 0.5$, $Q = 0.8$. From top to bottom, the corresponding observer inclination angles are: 17°, 53°, °, and °, with the black hole mass set to $M = 1$

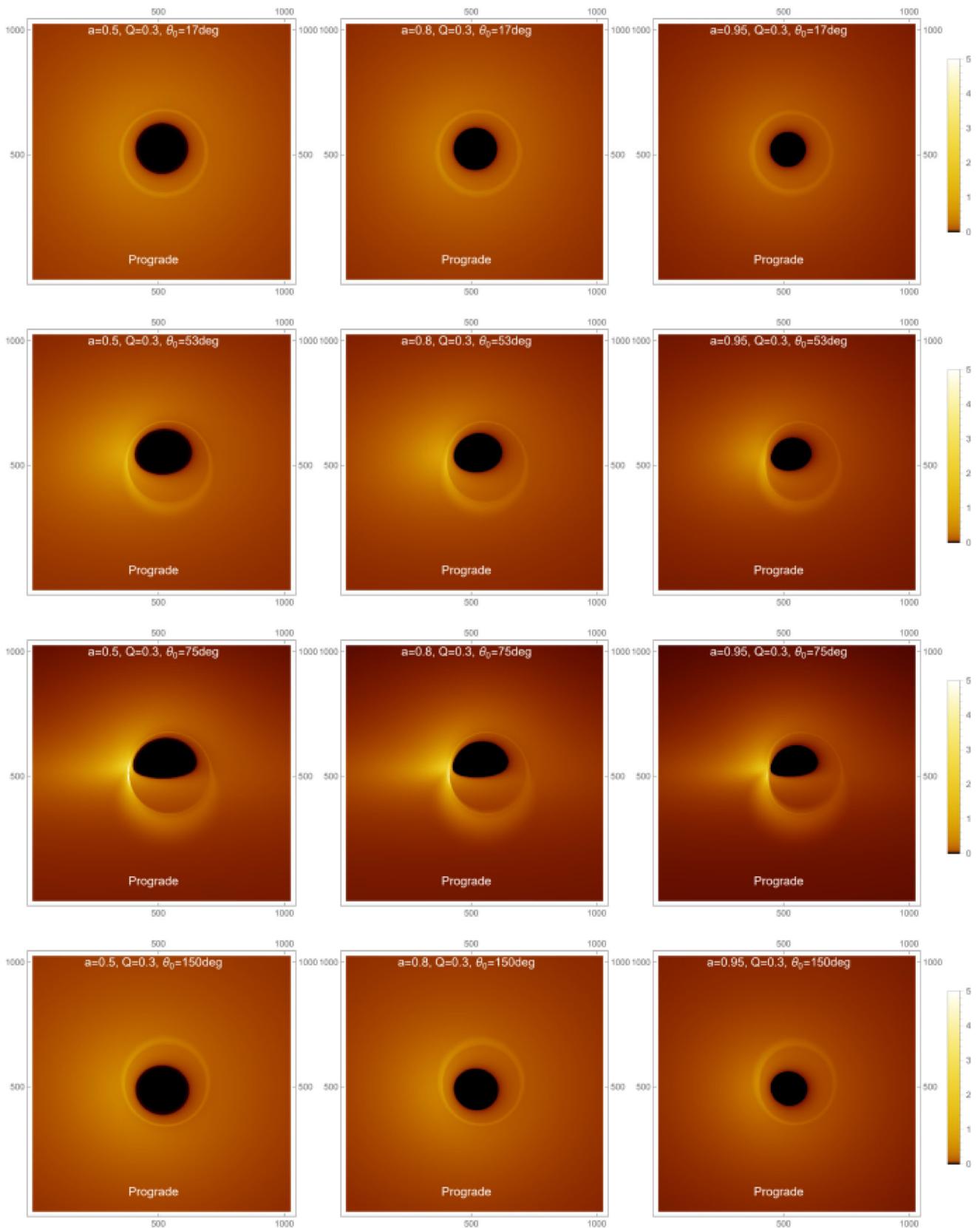

**Fig. 28** Images of the Kerr–Sen black hole surrounded by a prograde accretion disk at 86 GHz. From left to right, the parameters are: $a = 0.5$, $Q = 0.3$, $a = 0.8$, $Q = 0.3$, and $a = 0.95$, $Q = 0.3$. From top to bottom, the corresponding observer inclination angles are: 17°, °, 75°, and 150°, with the black hole mass set to $M = 1$

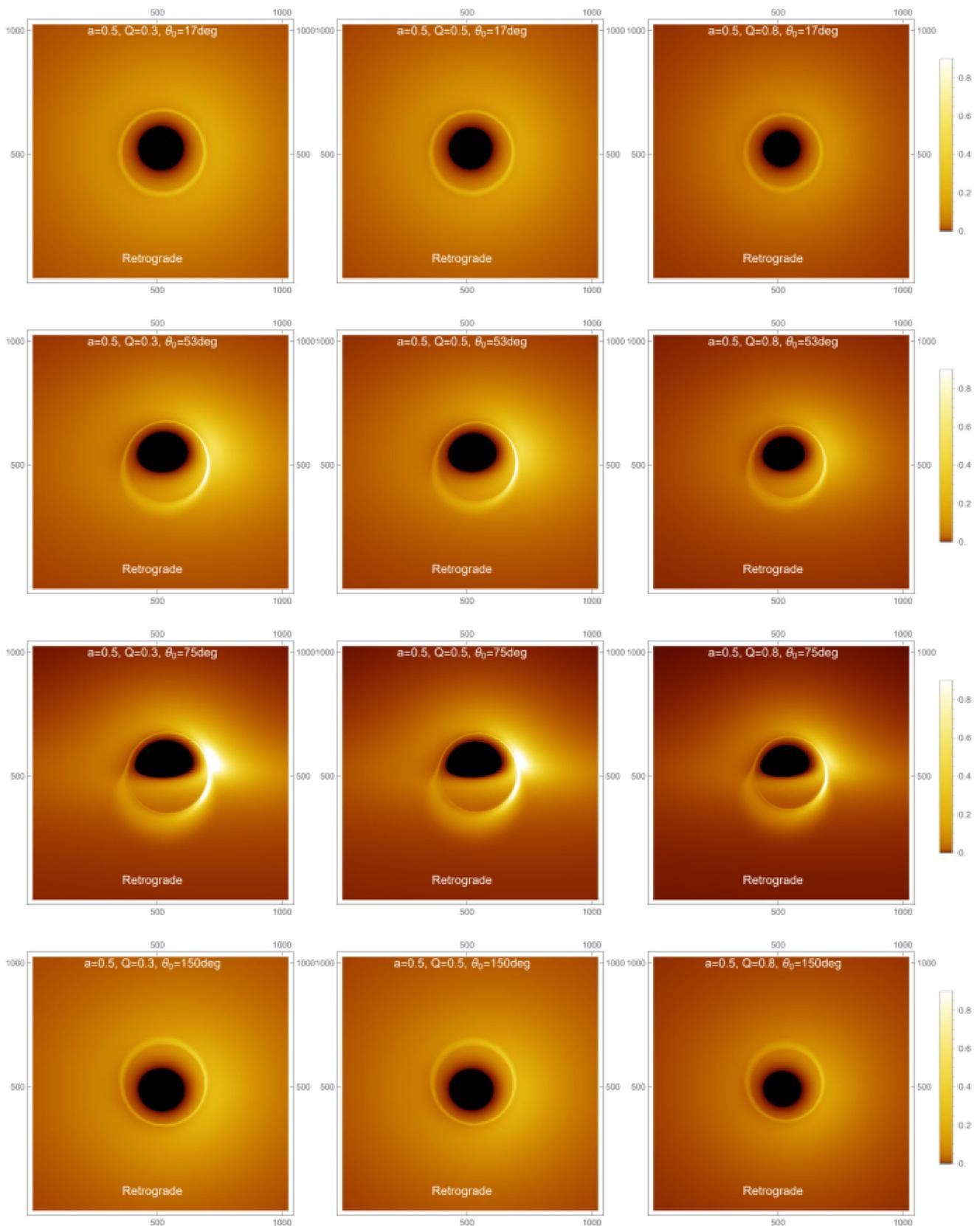

**Fig. 29** Images of the Kerr–Sen black hole surrounded by a retrograde accretion disk at 86 GHz. From left to right, the parameters are: $a = 0.5$, $Q = 0.3$, $a = 0.5$, $Q = 0.5$, and $a = 0.5$, $Q = 0.8$. From top to bottom, the corresponding observer inclination angles are: $17°$, $53°$, $75°$, and $150°$, with the black hole mass set to $M = 1$

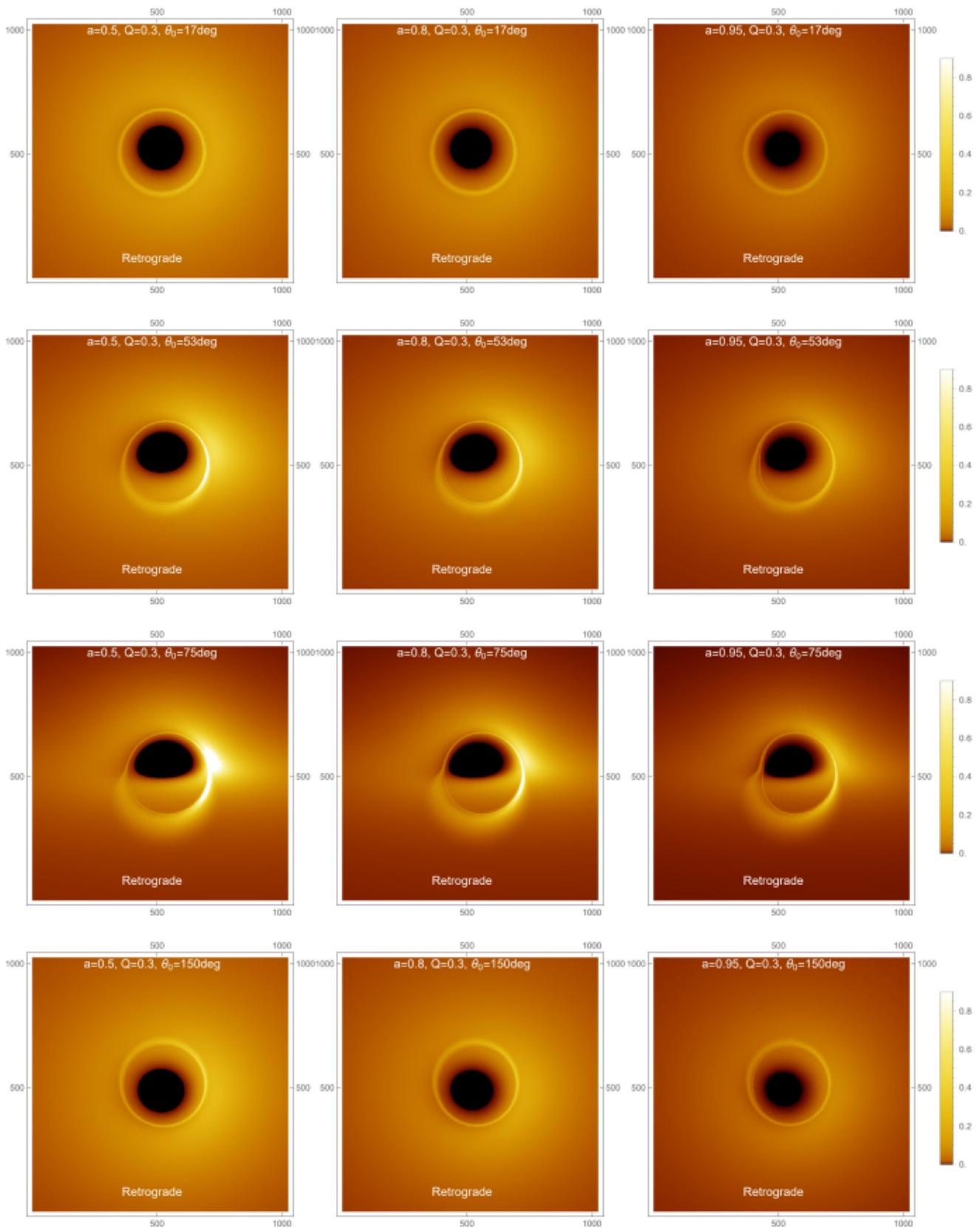

**Fig. 30** Images of the Kerr–Sen black hole surrounded by a retrograde accretion disk at 86 GHz. From left to right, the parameters are: $a = 0.5$, $Q = 0.3$, $a = 0.8$, $Q = 0.3$, and $a = 0.95$, $Q = 0.3$. From top to bottom, the corresponding observer inclination angles are: 17°, 53°, 75°, and 150°, with the black hole mass set to $M = 1$

and charge of the black hole significantly influence the distortion of the inner shadow, with the spin exerting a more pronounced effect than the charge.

Furthermore, we extended our analysis to investigate the impact of the observer's inclination angle on the observed intensity and redshift distribution. Our findings indicate that the inclination angle plays a pivotal role in shaping the redshift distribution, particularly near the innermost stable circular orbit (ISCO). As the observer's inclination angle increases, the redshift effect becomes more pronounced, with retrograde accretion disks exhibiting a stronger redshift, while prograde disks show blueshift on the left side and retrograde disks exhibit blueshift on the right. This study introduces an improved model for investigating redshift and light propagation around black holes, thereby enhancing our understanding of electromagnetic signatures from charged, rotating black holes.

Additionally, we performed a comparative analysis of the intensity distribution at different observational frequencies, specifically at 230 GHz and 86 GHz. The results demonstrate that both the peak intensity and total intensity at 86 GHz are significantly higher than those at 230 GHz. However, intrinsic spacetime features, such as the inner shadow and critical curve, remain invariant across both frequencies. The frequency dependence is of particular importance, as it provides deeper insights into the emission mechanisms of the accretion disk and the photon ring morphology, offering valuable perspectives for interpreting observational data from instruments such as the EHT.

Lastly, our results highlight the crucial role of the accretion disk's geometry in shaping the observable features of the black hole, including the photon ring and shadow. Our simulations of both prograde and retrograde accretion disks demonstrate asymmetries in the black hole image, which are modulated by the spin and charge of the black hole, as well as the observer's inclination angle. This research lays the groundwork for future observational investigations and opens new pathways for exploring the complex interactions between charged, rotating black holes and their accretion disks.

**Acknowledgements** This work is supported by the National Key RD Program of China (2024YFA1611700), the Guangxi Talent Program ("Highland of Innovation Talents") and the Fund Project of Chongqing Normal University (Grant Number: 24XLB033).

**Data Availability Statement** This manuscript has associated data in a data repository. [Authors' comment: All the data are shown as the figures and formulae in this paper. No other associated movie or animation data.]

**Code Availability Statement** This manuscript has associated code/software in a data repository. [Authors' comment: Code/Software sharing not applicable to this article as no code/software was generated or analysed during the current study.]

## References

1. C. Bambi, Testing black hole candidates with electromagnetic radiation. Rev. Mod. Phys. **89**, 025001 (2017). https://doi.org/10.1103/RevModPhys.89.025001
2. B.P. Abbott et al. (LIGO Scientific, Virgo), Observation of gravitational waves from a binary black hole merger. Phys. Rev. Lett. **116**, 061102 (2016). https://doi.org/10.1103/PhysRevLett.116.061102. arXiv:1602.03837 [gr-qc]
3. K. Akiyama et al. (Event Horizon Telescope), First M87 event horizon telescope results. I. The shadow of the supermassive black hole. Astrophys. J. Lett. **875**, L1 (2019). https://doi.org/10.3847/2041-8213/ab0ec7. arXiv:1906.11238 [astro-ph.GA]
4. K. Akiyama et al. (Event Horizon Telescope), First M87 event horizon telescope results. VII. Polarization of the ring. Astrophys. J. Lett. **910**, L12 (2021). https://doi.org/10.3847/2041-8213/abe71d. arXiv:2105.01169 [astro-ph.HE]
5. K. Akiyama et al. (Event Horizon Telescope), First M87 event horizon telescope results. VIII. Magnetic field structure near the event horizon. Astrophys. J. Lett. **910**, L13 (2021). https://doi.org/10.3847/2041-8213/abe4de. arXiv:2105.01173 [astro-ph.HE]
6. K. Akiyama et al. (Event Horizon Telescope), First Sagittarius A* event horizon telescope results. I. The shadow of the supermassive black hole in the center of the milky way. Astrophys. J. Lett. **930**, L12 (2022). https://doi.org/10.3847/2041-8213/ac6674. arXiv:2311.08680 [astro-ph.HE]
7. S. Chen, J. Jing, W.-L. Qian, B. Wang, Black hole images: a review. Sci. China Phys. Mech. Astron. **66**, 260401 (2023). https://doi.org/10.1007/s11433-022-2059-5. arXiv:2301.00113 [astro-ph.HE]
8. S. Chen, J. Jing, Kerr black hole shadows from axion-photon coupling. JCAP **05**, 023 (2024). https://doi.org/10.1088/1475-7516/2024/05/023. arXiv:2310.06490 [gr-qc]
9. Y. Hou, P. Liu, M. Guo, H. Yan, B. Chen, Multi-level images around Kerr–Newman black holes. Class. Quantum Gravity **39**, 194001 (2022). https://doi.org/10.1088/1361-6382/ac8860. arXiv:2203.02755 [gr-qc]
10. Q. Sun, Y. Zhang, C.-H. Xie, Q.-Q. Li, Shadow of Kerr black hole surrounded by a cloud of strings in Rastall gravity and constraints from M87*. Phys. Dark Univ. **46**, 101599 (2024). https://doi.org/10.1016/j.dark.2024.101599. arXiv:2401.08693 [gr-qc]
11. V. Bozza, Gravitational lensing by black holes. Gen. Relativ. Gravit. **42**, 2269 (2010). https://doi.org/10.1007/s10714-010-0988-2. arXiv:0911.2187 [gr-qc]
12. H. Falcke, F. Melia, E. Agol, Viewing the shadow of the black hole at the galactic center. Astrophys. J. Lett. **528**, L13 (2000). https://doi.org/10.1086/312423. arXiv:astro-ph/9912263
13. Y. Meng, X.-M. Kuang, X.-J. Wang, B. Wang, J.-P. Wu, Images of hairy Reissner–Nordström black hole illuminated by static accre-

tions. Eur. Phys. J. C **84**, 305 (2024). https://doi.org/10.1140/epjc/s10052-024-12686-w. arXiv:2401.05634 [gr-qc]

14. Y. Hou, Z. Zhang, H. Yan, M. Guo, B. Chen, Image of a Kerr–Melvin black hole with a thin accretion disk. Phys. Rev. D **106**, 064058 (2022). https://doi.org/10.1103/PhysRevD.106.064058. arXiv:2206.13744 [gr-qc]

15. T. Johannsen, A.E. Broderick, P.M. Plewa, S. Chatzopoulos, S.S. Doeleman, F. Eisenhauer, V.L. Fish, R. Genzel, O. Gerhard, M.D. Johnson, Testing general relativity with the shadow size of Sgr A*. Phys. Rev. Lett. **116**, 031101 (2016). https://doi.org/10.1103/PhysRevLett.116.031101. arXiv:1512.02640 [astro-ph.GA]

16. S.E. Gralla, A. Lupsasca, D.P. Marrone, The observer's guide to the Kerr–Newman black hole. Phys. Rev. D **101**, 044032 (2020). https://doi.org/10.1103/PhysRevD.101.044032. arXiv:1910.12873 [gr-qc]

17. S.E. Gralla, D.E. Holz, R.M. Wald, Black hole shadows, photon rings, and lensing rings. Phys. Rev. D **100**, 024018 (2019). https://doi.org/10.1103/PhysRevD.100.024018

18. A. Grenzebach, V. Perlick, C. Lämmerzahl, Photon regions and shadows of Kerr–Newman-NUT black holes with a cosmological constant. Phys. Rev. D **89**, 124004 (2014). https://doi.org/10.1103/PhysRevD.89.124004. arXiv:1403.5234 [gr-qc]

19. R. Narayan, M.D. Johnson, C.F. Gammie, The shadow of a spherically accreting black hole. Astrophys. J. Lett. **885**, L33 (2019). https://doi.org/10.3847/2041-8213/ab518c. arXiv:1910.02957 [astro-ph.HE]

20. M.A. Abramowicz, P.C. Fragile, Foundations of black hole accretion disk theory. Living Rev. Relativ. **16**, 1 (2013). https://doi.org/10.12942/lrr-2013-1. arXiv:1104.5499 [astro-ph.HE]

21. S.A. Balbus, J.F. Hawley, Instability, turbulence, and enhanced transport in accretion disks. Rev. Mod. Phys. **70**, 1 (1998). https://doi.org/10.1103/RevModPhys.70.1

22. A. Ricarte, D. Palumbo, R. Narayan, F. Roelofs, E. Emami, Observational signatures of frame dragging in strong gravity. Astrophys. J. Lett. **941**, L12 (2022). https://doi.org/10.3847/2041-8213/aca087. arXiv:2211.01810 [gr-qc]

23. K. Akiyama et al. (Event Horizon Telescope), The polarized image of a synchrotron-emitting ring of gas orbiting a black hole. Astrophys. J. **912**, 35 (2021). https://doi.org/10.3847/1538-4357/abf117. arXiv:2105.01804 [astro-ph.HE]

24. H. Paugnat, A. Lupsasca, F. Vincent, M. Wielgus, Photon ring test of the Kerr hypothesis: variation in the ring shape. Astron. Astrophys. **668**, A11 (2022). https://doi.org/10.1051/0004-6361/202244216. arXiv:2206.02781 [astro-ph.HE]

25. K. Wang, C.-J. Feng, T. Wang, Image of Kerr-de Sitter black holes illuminated by equatorial thin accretion disks. Eur. Phys. J. C **84**, 457 (2024). https://doi.org/10.1140/epjc/s10052-024-12825-3. arXiv:2309.16944 [gr-qc]

26. P. Hintz, A. Vasy, The global non-linear stability of the Kerr–de Sitter family of black holes. Commun. Math. Phys. **350**, 913 (2017). https://doi.org/10.1007/s00220-017-3015-6. arXiv:1606.04014 [gr-qc]

27. V. Perlick, O.Y. Tsupko, Light propagation in a plasma on Kerr spacetime: separation of the Hamilton–Jacobi equation and calculation of shadows. Phys. Rev. D **92**, 104031 (2015). https://doi.org/10.1103/PhysRevD.92.104031. arXiv:1507.04217 [gr-qc]

28. S. Guo, Y.-X. Huang, E.-W. Liang, Y. Liang, Q.-Q. Jiang, K. Lin, Image of the Kerr–Newman black hole surrounded by a thin accretion disk. Astrophys. J. **975**, 237 (2024). https://doi.org/10.3847/1538-4357/ad7d85. arXiv:2411.07914 [astro-ph.HE]

29. H. Zhang, S. Chen, J. Jing, Photon ring and observational appearance of a charged spinning black hole. Phys. Rev. D **107**, 104042 (2023). https://doi.org/10.1103/PhysRevD.107.104042. arXiv:2303.11865 [gr-qc]

30. O.Y. Tsupko, G.S. Bisnovatyi-Kogan, Charged black hole shadows: analytic formulas for bending angle and image contours. Phys.

Rev. D **106**, 064033 (2022). https://doi.org/10.1103/PhysRevD.106.064033. arXiv:2205.00695

31. D. García, D. Galtsov, O. Kechkin, Stringy black hole shadows: lensing by Kerr–Sen and dyonic dilaton black holes. Phys. Rev. D **102**, 124058 (2020). https://doi.org/10.1103/PhysRevD.102.124058. arXiv:2008.05144 [hep-th]

32. R. Kumar, S.G. Ghosh, Photon ring structure and observational signatures of Kerr–Sen black holes. Eur. Phys. J. C **82**, 512 (2022). https://doi.org/10.1140/epjc/s10052-022-10413-x. arXiv:2201.03224 [gr-qc]

33. A. Sen, Rotating charged black hole solution in heterotic string theory. Phys. Rev. Lett. **69**, 1006 (1992). https://doi.org/10.1103/PhysRevLett.69.1006

34. Z. Younsi, O. Porth, Y. Mizuno, GRMHD simulations of magnetized accretion onto Kerr–Sen black holes. Mon. Not. R. Astron. Soc. **516**, 1143 (2022). https://doi.org/10.1093/mnras/stac1962. arXiv:2205.10174 [astro-ph.HE]

35. Y. Li, H. Yang, Y. Yuan, Probing stringy charges via black hole shadow deformations: a Kerr–Sen case study. JHEP **10**, 115 (2022). https://doi.org/10.1007/JHEP10115. arXiv:2206.01774 [hep-th]

36. A. Hioki, K. Suzuki, S. Tomizawa, Photon trajectories and shadows in Kerr–Sen spacetime. Phys. Rev. D **106**, 044022 (2022). https://doi.org/10.1103/PhysRevD.106.044022. arXiv:2203.13898 [gr-qc]

37. F.H. Vincent, M. Wielgus, E. Gourgoulhon, Geodesic motion and shadow of a Kerr-like black hole. Astron. Astrophys. **646**, A55 (2021). https://doi.org/10.1051/0004-6361/202140980. arXiv:2102.09256 [gr-qc]

38. E. Collaboration, First M87 event horizon telescope results VII polarization of the ring. Astrophys. J. Lett. **910**, L12 (2021). https://doi.org/10.3847/2041-8213/abf1c5. arXiv:2105.01169 [astro-ph.HE]

39. A.E. Broderick, R. Gold, M. Karami, Frequency-dependent black hole shadow and photon ring in the EHT band. Astrophys. J. **940**, 57 (2022). https://doi.org/10.3847/1538-4357/ac91cc. arXiv:2208.09004 [astro-ph.HE]

40. G.T. Horowitz, D.L. Welch, Exact three-dimensional black holes in string theory. Phys. Rev. Lett. **71**, 328 (1993). https://doi.org/10.1103/PhysRevLett.71.328

41. H. Feng, R.-J. Yang, W.-Q. Chen, Thin accretion disk and shadow of Kerr–Sen black hole in Einstein–Maxwell-dilaton-axion gravity. Astropart. Phys. **166**, 103075 (2025). https://doi.org/10.1016/j.astropartphys.2024.103075. arXiv:2403.18541 [gr-qc]

42. B. Carter, Global structure of the Kerr family of gravitational fields. Phys. Rev. **174**, 1559 (1968). https://doi.org/10.1103/PhysRev.174.1559

43. S. Chandrasekhar, *The Mathematical Theory of Black Holes* (Oxford University Press, Oxford, 1983)

44. E. Hackmann, V. Kagramanova, J. Kunz, C. Lämmerzahl, Analytic solutions of the geodesic equation in Kerr-(anti)de Sitter spacetimes. Phys. Rev. D **81**, 044020 (2010). https://doi.org/10.1103/PhysRevD.81.044020. arXiv:1009.6117 [gr-qc]

45. M. Wang, S. Chen, J. Jing, Chaotic motion of particles in the Kerr–Sen black hole spacetime. Phys. Rev. D **102**, 104021 (2020). https://doi.org/10.1103/PhysRevD.102.104021. arXiv:2010.00155 [gr-qc]

46. S.K. Sahoo, N. Yadav, I. Banerjee, Imprints of Einstein–Maxwell-dilaton-axion gravity in the observed shadows of Sgr A* and M87*. Phys. Rev. D **109**, 044008 (2024). https://doi.org/10.1103/PhysRevD.109.044008. arXiv:2305.14870 [gr-qc]

47. S.E. Gralla, A. Lupsasca, Null geodesics of the Kerr exterior. Phys. Rev. D **101**, 044032 (2020). https://doi.org/10.1103/PhysRevD.101.044032. arXiv:1910.12881 [gr-qc]

48. H. Yang, H. Zhang, Photon trajectories in Kerr–Sen spacetime: a mino time approach. Phys. Rev. D **104**, 104053 (2021). https://doi.org/10.1103/PhysRevD.104.104053. arXiv:2106.04218 [gr-qc]

49. D. Kapec, A. Lupsasca, Universal Teukolsky equations and black hole photon rings. JHEP **09**, 134 (2021). https://doi.org/10.1007/JHEP09134. arXiv:2106.05276 [hep-th]

50. S.E. Gralla, A. Lupsasca, Lensing by Kerr black holes. Phys. Rev. D **101**, 044031 (2020). https://doi.org/10.1103/PhysRevD.101.044031. arXiv:1910.12873 [gr-qc]

51. B.-R. Chen, T. Hsieh, D.-S. Lee, Null geodesics in extremal Kerr–Newman black holes. Phys. Rev. D **111**, 024058 (2025). https://doi.org/10.1103/PhysRevD.111.024058. arXiv:2411.00458 [gr-qc]

52. A. Chael, M.D. Johnson, A. Lupsasca, Observing the inner shadow of a black hole: a direct view of the event horizon. Astrophys. J. **918**, 6 (2021)

53. K. Akiyama et al. (Event Horizon Telescope), First M87 event horizon telescope results. III. Data processing and calibration. Astrophys. J. Lett. **875**, L3 (2019). https://doi.org/10.3847/2041-8213/ab0c57. arXiv:1906.11240 [astro-ph.GA]